\documentclass[preprint,12pt]{JHEP3} % 10pt is ignored!
\usepackage{epsfig,multicol}

\newcommand\fverb{\setbox\varpippobox=\hbox\bgroup\verb}
\newcommand\fverbdo{\egroup\medskip\noindent%
			\fbox{\unhbox\varpippobox}\ }
\newcommand\fverbit{\egroup\item[\fbox{\unhbox\varpippobox}]}
\newbox\varpippobox
\newcommand{\EQ}{\begin{equation}}
\newcommand{\EN}{\end{equation}}
\newcommand{\bea}{\begin{eqnarray}}
\newcommand{\ena}{\end{eqnarray}}
\newcommand{\bdis}{\begin{displaymath}}
\newcommand{\edis}{\end{displaymath}}

\renewcommand{\a}{\alpha}
\renewcommand{\b}{\beta}

\renewcommand{\t}{\tau}

\newcommand{\pa}{\partial}

\newcommand{\nn}{\nonumber \\}

\title{Background Field Method in Stochastic Quantization of  ${\cal N}$ = 1 Supersymmetric Yang-Mills Theory}

\author{Naohito\ Nakazawa\\
     High Energy Accelerator Research Organization (KEK) \\
     Tsukuba 305-0801, Japan\\
	 E-mail: \email{naohito@post.kek.jp}}
\received{February 20, 2001} 		%%
\revised{May 1, 2001}
\accepted{November 27, 2001}		%% These are for published papers.

\abstract{In the previous works, we proposed the stochastic quantization method (SQM) approach to ${\cal N}=1$ supersymmetric Yang-Mills theory (SYM). In four dimensions, in particular, we obtained the superfield Langevin equation and the corresponding Fokker-Planck equation which describe the underlying stochastic process manifestly preserving the global supersymmetry as well as the local gauge symmetry.  The stochastic gauge fixing procedure was also applied to SYM$_4$ in the superfield formalism. In this note, we apply the background field methd to SYM$_4$ in terms of the stochastic action principle in the SQM approach. The one-loop $\beta$-function for the gauge coupling agrees with that given by the path-integral approach, thereby confirming that the stochastic gauge fixing procedure with the background local gauge invariant Zwanziger's gauge fixing functions simulates the contributions from the Nielsen-Kallosh ghost as well as the Faddeev-Popov ghost at the one-loop level. We also show the equivalence of the effective stochastic action in the background field method to the standard one in SQM.}

\begin{document} 

\newcommand{\caldslash}{{\cal D}\!\!\!\! \slash}

%\maketitle  IS IGNORED %%%%%%%%%%%

%%%%%%%%%%%%%%%%%%%%%%%%%%%%%%%%%%%%%%%%%%%%%%%%%%%%%%%%%%%%%%%%%%%%%%%%%%%%%%
%\section{Introduction}                   2003.2.10 version 
%%%%%%%%%%%%%%%%%%%%%%%%%%%%%%%%%%%%%%%%%%%%%%%%%%%%%%%%%%%%%%%%%%%%%%%%%%%%%%
\section{Introduction}

In the analysis of gauge theories, the background field method (BFM) simplifies the perturbative calculation and the renormalization procedure by respecting the background local gauge invariance\cite{D-H}\cite{'tHooft}\cite{Abbott}\cite{IO-CM-JO}\cite{Boulware}\cite{Hart}\cite{AGS}. The method is, in particular, a powerful tool for supersymmetric models in the superfield formalism\cite{HSKS}\cite{GSR}\cite{GS}\cite{GZ}. In the previous works\cite{Nakazawa1}\cite{Nakazawa2}, we proposed an approach to ${\cal N}=1$ supersymmetric Yang-Mills theory (SYM)\cite{FZ} via the stochastic quantization method (SQM)\cite{PW}. The superfield Langevin equation and the corresponding Fokker-Planck equation describe the underlying stochastic process for SYM$_4$ manifestly preserving the global supersymmetry as well as the local gauge symmetry\cite{Nakazawa1}. It is the extension of prior works\cite{BGZ}\cite{Ishikawa}\cite{Kalivas} on the application of SQM to the supersymmetric U(1) gauge theory\cite{WZ1} to SU(N) case. The stochastic gauge fixing procedure for Yang-Mills theory (YM)\cite{Zwanziger}\cite{BZ}\cite{NOOY}\cite{BHST} was also applied to SYM$_4$ in the superfield formalism\cite{Nakazawa2}. In Ref.\cite{Nakazawa2}, it was shown that the stochastic gauge fixing procedure for SYM$_4$ is equivalent to the Faddeev-Popov prescription in the path-integral method. The renormalizability of SYM$_4$ in the SQM approach is ensured in terms of the BRST invariant stochastic action principle. For further application of SQM to SYM, such as the evaluation of anomalies\cite{AGS-C}\cite{CPS}\cite{PiSi}\cite{GW-GMZ}, we fomulate BFM in this context. In the SQM approach, BFM has been applied to YM$_4$\cite{Okano}. The equivalence of the stochastic gauge fixing procedure to the Faddeev-Popov prescription has been established by the explicit calculation up to the two loop order of the perturbation\cite{MOSZ}. In SYM$_4$, in contrast to the standard path-integral for SYM$_4$\cite{FP}, the Faddeev-Popov prescription in the conventional BFM requires an extra ghost\cite{GSR}, a Nielsen-Kallosh type ghost\cite{N-K}, in addition to the ordinary Faddeev-Popov ghost. The reason is that the chiral condition on the Nakanishi-Lautrup superfield and the Faddeev-Popov ghost superfield is replaced to the background chiral condition in BFM. This causes a non-trivial contribution of the Nakanishi-Lautrup field. The additional Nielsen-Kallosh ghost is introduced to cancel the contribution\cite{GSR}. Therefore, it is non-trivial in BFM formulated in the SQM approach whether the stochastic gauge fixing procedure simulates the contributions from both these ghosts. In this note, we apply BFM to SYM$_4$ in the context of SQM. We show that the one-loop $\beta$-function for the gauge coupling obtained by BFM in SQM agrees with the well-known result in ${\cal N}=1$ SYM$_4$\cite{FZ}. This indicates that appropriate Zwanziger's gauge fixing functions, which are chosen to be invariant under the background local gauge transformation, simulates the contributions from the Nielsen-Kallosh ghost as well as the Faddeev-Popov ghost. The paper is organized as follows. In \S 2, we recapitulate the stochastic gauge fixing procedure for SYM$_4$. In \S 3, the stochastic action principle\cite{Gozzi}\cite{NY-CKS}\cite{CH}\cite{ZZ}\cite{KOT}\cite{Nakazawa3} is introduced for the perturbative analysis of SYM$_4$. In \S 4, we apply BFM to SYM$_4$ in SQM. A slightly improved point in comparison with the YM case\cite{Okano}\cite{MOSZ} is that we introduce a background superfield for the auxiliary superfield, which is the canonical conjugate momentum of the vector superfield with respect to the stochastic time, in order to define a  background local gauge invariant \lq\lq classical \rq\rq stochastic action. In \S 5, the $\beta$-functions are determined in the one-loop level. We use the regularization via the dimensional reduction.\cite{Siegel}\cite{CJN} In \S 6, we show the equivalence of an effective stochastic action defined by the stochastic BFM, which is a generator of the 1-P-I stochastic vertices, to the standard one in SQM. This explains the reason why BFM in SQM reproduces the contribution of the Nielsen-Kallosh ghost as well as the Faddeev-Popov ghost. 
In \S 7, we discuss the BRST invariant formulation of SQM. In particular, we complete the proof in \S 6 on the equivalence of the stochastic BFM to the standard SQM in terms of their BRST invariant formulations. 
%%%%%%%%%%%%%%%%%%%%%%%%%%%%%%%%%%%%%%%%%%%%%%%%%%%%%%%%%%%%%%%%%%%%%%%%%%%%%%
%The proof of the equivalence is completed by showing that the background field
%dependence of the expectation values evaluated in the stochastic BFM, which can% be mapped to those defined in the standard SQM with an unusual background 
%dependent stochastic gauge fixing, is at most a BRST exact term. 
%%%%%%%%%%%%%%%%%%%%%%%%%%%%%%%%%%%%%%%%%%%%%%%%%%%%%%%%%%%%%%%%%%%%%%%%%%%%%%
Section 8 is devoted to discussions. In Appendix A, the conventions on BFM is explained. The Langevin equation for BFM is derived in Appendix B.  Throughout this note, we call BFM in the SQM approach as \lq\lq stochastic BFM \rq\rq. While we refer to BFM in the path-integral approach as \lq\lq conventional BFM \rq\rq. The adjective \lq\lq standard \rq\rq means that both approaches, SQM and the path-integral approach, are supposed without BFM.

%\newpage
%%%%%%%%%%%%%%%%%%%%%%%%%%%%%%%%%%%%%%%%%%%%%%%%%%%%%%%%%%%%%%%%%%%%%%%%%%%%%%%
%section{Stochastic gauge fixing for SYM$_4$}
%%%%%%%%%%%%%%%%%%%%%%%%%%%%%%%%%%%%%%%%%%%%%%%%%%%%%%%%%%%%%%%%%%%%%%%%%%%%%%%

%
\section{Stochastic gauge fixing for SYM$_4$}

In the superfield formalism, without choosing the Wess-Zumino gauge, the action for SYM$_4$ is non-polynomial. 
\bea
\label{eq:SYM-4Daction}
S 
 =  
- \int\!\!\! d^4x d^2\theta d^2{\bar \theta} \displaystyle{\frac{1}{4 g^2}} {\rm Tr}    \Bigl(
W^\alpha W_\alpha \delta^2 ( {\bar \theta} ) + {\overline W}_{\dot \alpha}{\overline W}^{\dot \alpha} \delta^2 ( \theta )
\Bigr)           \  . 
\ena
Here 
\bea
\label{eq:Gluino-field}
W_\alpha 
 = &
- \displaystyle{\frac{1}{8}} {\overline D}^2 {\rm e}^{-2gV}D_\alpha {\rm e}^{2gV} 
     \ , \qquad 
{\overline W}_{\dot \alpha} 
=  
 \displaystyle{\frac{1}{8}} D^2 {\rm e}^{2gV}{\overline D}_{\dot \alpha} {\rm e}^{-2gV}            \  .
\ena
We use a notation closely related with that in Ref.\cite{WB} by the analytic continuation to the Euclidean space-time, $i x^0 \equiv x^4$. The superfield $V$ is $SU(N)$ algebra-valued, $V \in su(N)$ ; $V = V^a t_a$, $[ t_a, t_b] = if_{abc} t_c$ and ${\rm tr}( t_a t_b ) = {\rm k}\delta_{ab}$.  For simplicity we use ${\rm Tr} \equiv \displaystyle{\frac{1}{\rm k}}{\rm tr}$.

In the non-polynomial form of the action (\ref{eq:SYM-4Daction}) in the superfield formalism, 
in particular, our proposal is presented in an analogy to LGT$_d$ in the context of SQM\cite{Nakazawa1}\cite{Nakazawa2}. For lattice gauge theories, SQM approach is well-established.\cite{DDH-GL-H-S-N} Therefore, SQM is applied to SYM$_4$ with the following dictionary. 
\bea
\label{eq:SYM-LGT-correspondence}
& {} & 
{\rm dynamical\ field}  :\ 
U_\mu \in U(N) ,\ {\rm link\ variable}    \nn
& & \qquad \qquad \leftrightarrow 
U \equiv e^{2gV},\  {\rm vector\ superfield}\ V,\  V \in su(N)               \ .   \nn  
& {} & 
{\rm local\ gauge\ transf.} : 
U_\mu 
\rightarrow  
e^{i \Lambda (x)} U_\mu e^{-i\Lambda (x+\mu )}      \nn
& &  \qquad \qquad \leftrightarrow  
U 
\rightarrow 
e^{-ig \Lambda^\dagger} U e^{ig \Lambda},\       
{\rm chiral\ superfields}\ {\overline D}_{\dot \a} \Lambda = D_\a \Lambda^\dagger = 0                       \ . \nn 
& {} & 
{\rm time\ development} :\  
( \delta U_\mu ) U_\mu^\dagger ,\ {\rm Maurer\!\!-\!\!Cartan\ form}  \nn
& &  \qquad \qquad  \leftrightarrow  
( \delta U ) U^{-1}         \ . \nn 
& {} & 
{\rm differential\ operator}  :\  
E_a, {\rm (left)\ Lie\ derivative}     ,\ E_a U_\mu= t_a U_\mu   \nn
& &  \qquad \qquad  \leftrightarrow  
{\hat {\cal E}}_a, \   {\rm analogue\ of\ the\ (left)\ Lie\ derivative}               \ . \nn 
& {} & 
{\rm integration\ measure}  : 
dU_\mu,\ {\rm Haar\ measure\ on}\  U(N)            \nn
& &  \qquad \qquad  \leftrightarrow 
\sqrt{G}{\cal D}V, \ {\rm analogue\ of\ the\ Haar\ measure}     \ . \nonumber
\nonumber
\ena

Although $U \equiv e^{2gV}$ is not a group element, the differential operator ${\hat {\cal E}}_a$
\bea
\label{eq:Lie-derivative-analogue-eq1}
{\hat {\cal E}}_a (z) U (z') 
= 
t_a U (z') \delta^8( z-z' )      \ , 
\ena
 and the path-integral measure $\sqrt{G}{\cal D}V$ are constructed in an analogous manner as those on a group manifold. Here  
 $z \equiv ( x^m, \theta_\a, {\bar \theta}_{\dot \a} )$ 
 denotes the superspace coordinates. 
 Thanks to the measure introduced with a non-trivial metric, the partial integration with respect to the differential operator holds and the Langevin equation can be translated to the corresponding Fokker-Planck equation. 
The superfield Langevin equation for SYM$_4$ is derived by regarding $e^{2gV}$, instead of $V$, as a fundamental variable\cite{Nakazawa1}  
\bea
\label{eq:Langevin-eq1}
\displaystyle{\frac{1}{2g}}
\left( \Delta e^{2gV}  
   \right) e^{-2gV} ( \t, z )
& = & 
- \beta\Delta \t 2g{\hat {\cal E}} ( \t, z ) S + \Delta w ( \t, z )       \ , \nn
\langle \Delta w_{ij} ( \t, z ) \Delta w_{kl} ( \t, z' ) 
\rangle_{\Delta w_\t}   
& = & 
  \beta\Delta \t 2{\rm k}\Bigl( 
\delta_{il}\delta_{jk} - \displaystyle{\frac{1}{N}}\delta_{ij}\delta_{kl} 
\Bigr) \delta^8 ( z - z' )        \ .
\ena
Here 
${\hat {\cal E}} ( \t, z ) \equiv t_a {\hat {\cal E}}_a ( \t, z ) $, and 
$
{\hat {\cal E}} ( \t, z ) S 
= 
{1\over 4g^2} \left(    
e^{2g L_V} {\cal D}^\a W_\a + {\overline {\cal D}}_{\dot \a} {\overline W}^{\dot \a}      \right) \ . 
$ 
The operator $L_V$ is defined by 
$L_V X \equiv [\ V,\ X\ ]$ for X in the adjoint representation.\cite{GGRS} 
The time evolution is described in a discretized notation with respect to the stochastic time $\t$ to allow a clear understanding. 
$\langle ... \rangle_{\Delta w_\t}$ denotes that the expectation value is evaluated by means of the noise correlation at the stochastic time $\t$. $\beta$ is introduced as a scaling parameter of the stochastic time, which is necessary for the multiplicative renormalization in SQM.  
The covariant spinor derivatives are defined by 
$
{\cal D}_\a \equiv e^{-2gV} D_\a e^{2gV} 
$ and 
$
{\overline {\cal D}}_{\dot \a} \equiv e^{2gV} {\overline D}_{\dot \a} e^{-2gV} 
$. 
The operations of these covariant spinor derivatives in the adjoint representation are defined as the commutation or anti-commutation relations. This means, for example, that 
the equations of motion are understood as  
$
{\cal D}^\a W_\a \equiv \Big\{ {\cal D}^\a, W_\a \Big\}  ,\ 
$
$
{\overline {\cal D}}_{\dot \a} {\overline W}^{\dot \a} \equiv \Big\{ {\overline {\cal D}}_{\dot \a}, {\overline W}^{\dot \a} \Big\}  .\ 
$ 
We note that the reality condition\cite{GSR} implies 
$e^{2g L_V} {\cal D}^\a W_\a = {\overline {\cal D}}_{\dot \a}{\overline W}^{\dot \a}$.
Therefore, we discard one of them in (\ref{eq:Langevin-eq1}) in the perturbative calculation. 

By translating the variation $\Delta e^{2gV}$ to $\Delta V$, the Langevin equation (\ref{eq:Langevin-eq1}) reads
\bea
\label{eq:Langevin-eq1'}
\Delta V ( \t, z )
& = &
- \beta \Delta \t 
 \displaystyle{\frac{2g L_V}{{\rm e}^{2g L_V} - 1}} 
 \displaystyle{\frac{2g L_V}{1 - {\rm e}^{-2g L_V}}} 
 \displaystyle{\frac{\delta S}{\delta V^t}}   
+ \Delta_w \Xi ( \t, z )        \ , \nn 
\Delta_w \Xi ( \t, z ) 
& \equiv & 
\displaystyle{\frac{2g L_V}{{\rm e}^{2g L_V} - 1}} \Delta w ( \t , z )  \ . 
\ena
$\Delta_w \Xi$ is a collective noise superfield.
This means that we choose the superfield kernel 
$ ({\rm e}^{2g L_V} - 1 )^{-1} 2g L_V ( 1 - {\rm e}^{-2g L_V} )^{-1} 2g L_V  $ 
for the superfield Langevin equation.

The extended (\lq\lq extended \rq\rq means stochastic time $dependent$) local gauge invariance is defined as follows. We first introduce auxiliary superfields, $\Lambda$ and $\Lambda^\dagger$, which are chiral and anti-chiral respectively, into the Langevin equation by the (inverse) local gauge transformation,
$e^{2gV}
\rightarrow 
e^{ig \Lambda^\dagger} e^{2gV} e^{-ig \Lambda}  $. 
Then we redefine the auxiliary superfields as   
\bea
\label{eq:auxiliary-field-eq1}
\Phi 
\equiv 
\displaystyle{\frac{1}{\beta}}
\displaystyle{\frac{i}{g}} \displaystyle{\frac{e^{ - ig L_{\Lambda}} - 1}{L_{\Lambda}}} 
{\dot  \Lambda } \ , \
{\bar \Phi} 
\equiv 
\displaystyle{\frac{1}{\beta}}
\displaystyle{\frac{i}{g}} \displaystyle{\frac{e^{ - ig L_{\Lambda^\dagger}} - 1}{L_{\Lambda^\dagger}}} 
{\dot \Lambda}^\dagger      \ . 
\ena
Here we denote 
$
{\dot \Lambda} 
\equiv \displaystyle{\frac{\Delta \Lambda}{\Delta \t}}
$. 
$\Phi$ and ${\bar \Phi}$ are also chiral and anti-chiral, 
${\overline D}\Phi = D{\bar \Phi} = 0$, respectively. 
Then we obtain
\bea
\label{eq:Langevin-eq2} 
& {} & 
\displaystyle{\frac{1}{2g}} 
\left( \Delta e^{2gV} 
\right) e^{-2gV}  
+ \beta\Delta \t \displaystyle{\frac{i}{2}} \left( 
{\bar \Phi} - e^{2g L_V} \Phi \right)           \nn
& {} & \qquad\qquad\qquad =
- 
\beta\Delta\t \displaystyle{\frac{1}{2g}}  \left( 
e^{2g L_V} {\cal D}^\a W_\a + {\overline{\cal D}}_{\dot \a}{\overline W}^{\dot \a}  \right) 
+ \Delta w       \ . 
\ena 
The Langevin equation (\ref{eq:Langevin-eq2}) is covariantly transformed under the extended local gauge transformation 
\bea
\label{eq:extended-local-gauge-transf-eq1}
e^{2gV ( \t, z )} 
& \rightarrow & e^{-i g \Sigma^\dagger ( \t, z )}e^{2gV ( \t, z )}e^{ig \Sigma ( \t, z )}   \ ,  \nn
e^{ig \Lambda ( \t, z ) }        
& \rightarrow &
e^{ig \Lambda ( \t, z ) }  e^{ig \Sigma ( \t, z ) }    \ .
\ena
In particular, the auxiliary superfields are transformed as 
\bea
\label{eq:extended-local-gauge-transf-eq2}
\Phi 
& \rightarrow & 
\displaystyle{\frac{1}{\beta}}
 \displaystyle{\frac{i}{g}} \displaystyle{\frac{e^{- ig L_{\Sigma}} - 1}{L_{\Sigma}}} 
{\dot \Sigma}  
+ e^{- ig L_\Sigma} \Phi   \ , \nn
{\bar \Phi} 
& \rightarrow & 
\displaystyle{\frac{1}{\beta}}
 \displaystyle{\frac{i}{g}} \displaystyle{\frac{e^{ -ig L_{\Sigma_\dagger}} - 1}{L_{\Sigma^\dagger}}} 
{\dot \Sigma}^\dagger  
+ e^{-ig L_{\Sigma^\dagger}} {\bar \Phi}   \ .   
\ena
In a weak coupling region, (\ref{eq:extended-local-gauge-transf-eq2}) reads, 
$
\Phi \rightarrow 
\Phi + \displaystyle{\frac{1}{\beta}}{\dot \Sigma} + ig [ \Phi,\ \Sigma ]  \ . 
$
This indicates that the extended local gauge transformation is interpreted as a 5-dimensional local gauge transformation for which the chiral and anti-chiral superfields play a role of the \lq\lq 5-th\rq\rq component of the gauge field. We note that the stochastic time has a distinct meaning from other 4-dimensional space-time coordinates. Therefore the global supersymmetry is the 4-dimensional one. 

The probability distribution, obtained as a stationary solution of the Fokker-Planck equation in a specified stochastic gauge fixing is equivalent to the standard Faddeev-Popov distribution for SYM$_4$.\cite{Nakazawa2}  For the perturbative analysis, we choose the gauge fixing functions in such a way that the Langevin equation reproduces the superpropagator of the vector superfield defined in a one parameter family of covariant gauges in the path-integral approach. 
The simplest choices of the Zwanziger's gauge fixing functions, $\Phi$ and ${\bar \Phi}, $ are given by
\bea
\label{eq:stochastic-gauge-fixing-eq1}
\Phi 
 = i \displaystyle{\frac{\xi}{4}} 
 {\overline D}^2 D^2 V  , \ 
{\bar \Phi}
 = - i \displaystyle{\frac{\xi}{4}} 
 D^2 {\overline D}^2 V  \ . 
\ena
This is the supersymmetric extension of the Zwanziger gauge for YM theory. $\xi=1$ corresponds to the Feynman gauge.

%\newpage
%%%%%%%%%%%%%%%%%%%%%%%%%%%%%%%%%%%%%%%%%%%%%%%%%%%%%%%%%%%%%%%%%%%%%%%%%%%%%%%
%section{Stochastic action principle for SYM$_4$}
%%%%%%%%%%%%%%%%%%%%%%%%%%%%%%%%%%%%%%%%%%%%%%%%%%%%%%%%%%%%%%%%%%%%%%%%%%%%%%%

%
\section{Stochastic action principle}

For the perturbative approach in SQM, we introduce an action principle for SYM$_4$, the so-called stochastic action. We will apply BFM to this stochastic action and demonstrate an explicit one-loop calculation to obtain the $\beta$-function for the gauge coupling. The stochastic action is defined by the path-integral representation of the Langevin equation\cite{Gozzi}\cite{NY-CKS}\cite{CH}. 
Here we consider the continuum limit of the Langevin equation (\ref{eq:Langevin-eq2}) by taking $\Delta \t \rightarrow 0$. The noise superfield $\Delta w$ is replaced to $\eta$. The correlation is redefined by
\bea
\label{eq:continuum-noise-correlation-eq1}
\langle \eta_{ij} ( \t, z ) \eta_{kl} ( \t, z' ) 
\rangle_{\eta_\t}  
& = & 
 2{\rm k} \beta \Big( 
\delta_{il}\delta_{jk} - \displaystyle{\frac{1}{N}}\delta_{ij}\delta_{kl} 
\Big) \delta ( \t - \t' ) \delta^8 ( z - z' )        \ .
\ena
The integral representation of the noise correlation is given by 
\bea
\label{eq:integral-represent-eq1}
Z 
\equiv 
\int\!\!\! {\cal D}\eta {\rm exp}\Big( 
- \displaystyle{\frac{1}{4\beta}}
\!\! \int \!\!\! d^8z d\t {\rm Tr}  \eta ( z, \t )^2
\Big)           \ . 
\ena 
Then we insert an unity as a device for the Parisi-Sourlas type supersymmetry\cite{PS}. Supposing the Langevin equation of the form, $E (V) = \eta$, where $E (V)$ is defined by the continuum limit $\Delta \t \rightarrow 0$ of (\ref{eq:Langevin-eq2}), the device is written by 
\bea
\label{eq:Parisi-Sourlas-eq1}
1 
 =  
\int\!\!\! \sqrt{G}{\cal D}V 
\delta \Big( E(V) - \eta       
 \Big) {\rm det}\Big( \displaystyle{\frac{\delta E(V)}{\delta V}} \Big)  \ . 
%%%%%%%%%%%%%%%%%%%%%%%%%%%%%%%%%%%%%%%%%%%%%%%%%%%%%%%%%%%%%%%%%%%%%%%%%%%%%
%E (V)
%& \equiv & 
%\displaystyle{\frac{1}{2g}} 
%\displaystyle{\frac{d}{d\t}}( e^{2gV} ) e^{-2gV}     \nn
%& {} & 
%\quad
%+ \displaystyle{\frac{i}{2}} \beta  ( {\bar \Phi} - e^{2g L_V} \Phi ) 
%+ \displaystyle{\frac{1}{2 g}} \beta  ( 
%e^{2g L_V} {\cal D}^\a W_\a + {\overline{\cal D}}_{\dot \a}{\overline W}^{\dot %\a} )        
% \Big)  \ . 
%%%%%%%%%%%%%%%%%%%%%%%%%%%%%%%%%%%%%%%%%%%%%%%%%%%%%%%%%%%%%%%%%%%%%%%%%%%%%%
\ena
After the integral representation of the $\delta$-functional and integrating out the noise superfield $\eta$, we obtain 
\bea
\label{eq:Parisi-Sourlas-eq2}
Z 
\equiv 
\int\!\!\! \sqrt{G}{\cal D}V {\cal D}\Pi {\rm exp}\bigg( 
\int \!\!\! d^8z d\t {\rm Tr} \Bigl( 
- \beta \Pi^2 + i \Pi E(V) \Bigr) 
\bigg)  
 {\rm det}\Big( \displaystyle{\frac{\delta E(V)}{\delta V}} \Big)         \ . 
\ena 
Here $\Pi$ is an auxiliary superfield introduced for the integral representation of the $\delta$-functional.
The factor ${\rm det}\Big( \displaystyle{\frac{\delta E(V)}{\delta V}} \Big)$ is expressed as an effective action with auxiliary fermionic vector superfields, ${\overline \Psi}$ and $\Psi$, that appears to be non-polynomial. In the Feynman gauge, $\xi=1$, the effective action takes the form,  
\bea
\label{eq:Parisi-Sourlas-eq3}
\int \!\!\! d^8z d\t {\rm Tr} 
{\overline \Psi} \Big( 
\displaystyle{\frac{d}{d\t}} + 2\beta \square + 
O ( V )  \Big) \Psi      \ . 
\ena
This implies that the superpropagator of the auxiliary fermionic vecor superfields is a retarded one with respect to the stochastic time. Since it includes a step function $\theta ( \t -\t')$, the contribution may be evaluated with a prescription to specify $\theta (0) = \displaystyle{\frac{1}{2}} $. While (\ref{eq:Parisi-Sourlas-eq3}) also includes $\delta^8 (0)$ in its perturbative expansion with respect to $V$ which vanishes in the regularization via the dimensional reduction.\cite{Siegel}\cite{CJN} Therefore we may discard the contribution, ${\rm det}\Big( \displaystyle{\frac{\delta E(V)}{\delta V}} \Big)$ . 

The integral representation (\ref{eq:Parisi-Sourlas-eq2}) reads
\bea
\label{eq:invariant-stochastic-action-eq1}
Z 
& = &  
\int \sqrt{G}{\cal D}V{\cal D}\Pi e^{ K}        \ , \nn
K 
& \equiv & 
\!\!\! \int \!\!\! d^8z d\t 
{\rm Tr} \Big[ 
- \beta \Pi^2 
+i \Pi \Big\{  
\displaystyle{\frac{1}{2g}} \Bigl( 
\displaystyle{\frac{d}{d\t}} e^{2gV} \Bigr) e^{-2gV}                             \nn
& {} & 
\quad 
+ \displaystyle{\frac{i}{2}} \beta  \left( 
 {\bar \Phi} - e^{2g L_V} \Phi \right) 
+ \displaystyle{\frac{1}{2 g}} \beta  \left( 
e^{2g L_V} {\cal D}^\a W_\a + {\overline{\cal D}}_{\dot \a}{\overline W}^{\dot \a} \right) 
\Big\}  
\Big]    \ .  
\ena
For the extended local gauge invariance of the stochastic action (\ref{eq:invariant-stochastic-action-eq1}), 
the auxiliary superfield $\Pi$ is transformed  as 
$
\Pi \rightarrow e^{-ig \Sigma^\dagger} \Pi e^{ig \Sigma^\dagger}    \ . 
$
This means that $\Pi$ is not a vector superfield. By a field redefinition, 
$
\Pi = \displaystyle{\frac{2g L_V}{1- e^{-2g L_V}}}\varpi 
$,
we introduce a vector superfield, $\varpi = \varpi^\dagger$. The stochastic action (\ref{eq:invariant-stochastic-action-eq1}) is expressed as
\bea
\label{eq:invariant-stochastic-action-eq2}
Z 
& = &  
\int {\cal D}V{\cal D}\varpi e^{ K}        \ , \nn
K 
& \equiv & 
\!\!\! \int \!\!\! d^8z d\t 
 \Big[  
\displaystyle{\frac{1}{2\kappa}} G^{ab} \varpi_a \varpi_a 
+ i \varpi_a \Big\{ 
{\dot V}^a             
- \displaystyle{\frac{i}{4\kappa}}  ( {\bar \Phi}^b L_b^{\ a} -  L^a_{\ b}\Phi^b )            \nn
& {} & \qquad\qquad\qquad\qquad 
- \displaystyle{\frac{1}{4\kappa g}} 
\Big( 
L^a_{\ b} ( {\cal D}^\a W_\a )^b 
+ ( {\overline {\cal D}}_{\dot \a} {\overline W}^{\dot \a} )^b L_b^{\ a}
\Big)   
\Big\} 
\Big]   \ .     
\ena
Here we have redefined the scaling parameter $\beta \equiv - \displaystyle{\frac{1}{2\kappa}}$.\cite{Nakazawa2} $L_a^{\ b}$ is the inverse of an analogue of the Maurer-Cartan one-form coefficient defined by 
\bea
\label{eq:Maurer-Cartan-coefficient-eq1}
K_a^{\ b} (z)
& \equiv & 
 {\rm Tr}\Big( 
t_a \cdot \displaystyle{\frac{1 - e^{-2g L_{V(z)}}}{2g L_{V(z)}}} t^b  
\Big)              \ , \nn
L_a^{\ b} (z)
& \equiv & 
 {\rm Tr}\Big( 
t_a \cdot \displaystyle{\frac{2g L_{V(z)}}{1 - e^{-2g L_{V(z)}}}} t^b  
\Big)  \ .
\ena
$K_a^{\ c}L_c^{\ b} = L_a^{\ c}K_c^{\ b} = \delta_a^{\ b}$. 
 We have also introduced a metric $G_{ab} = K^c_{\ a}K^c_{\ b}$, $G^{ab} \equiv L_c^{\ a} L_c^{\ b}$ and $G \equiv {\rm det}G_{ab}$. 
The path-integral measure, $\sqrt{G}{\cal D}V$, is required for the change of the integration variable from $( \delta U ) U^{-1}$ to $\delta V$. By the change of the integration variable from $\Pi$ to $\varpi$, we also need $({\sqrt{G}})^{-1}{\cal D}\varpi$. Therefore the non-trivial measure cancels out. 

In the last of this section, we comment on the transformation property of the auxiliary field, $\varpi$, and the extended local gauge invariance of the stochastic action (\ref{eq:invariant-stochastic-action-eq2}). 
The auxiliary superfield, $\varpi$, is a vector superfield, while its transformation property is complicated. The infinitesimal form of the extended local gauge transformation is given by 
\bea
\label{eq:extended-local-gauge-transf-eq3}
\delta V^a 
& = & 
+ \displaystyle{\frac{i}{2}} L^a_{\ b} \Sigma^b 
- \displaystyle{\frac{i}{2}} {\Sigma^\dagger}^b L_b^{\ a}   \ , \nn 
\delta \varpi_a 
& = & 
- \displaystyle{\frac{i}{2}} \pa_a L^c_{\ b} \varpi_c \Sigma^b 
+ \displaystyle{\frac{i}{2}} {\Sigma^\dagger}^b \pa_a L_b^{\ c} \varpi_c  
             \ , \nn 
\delta \Phi^a 
& = & 
- \left( 2\kappa {\dot \Sigma}^a + g f^{abc} \Phi^b \Sigma^c \right)              \ , \nn        
\delta {\bar \Phi}^a 
& = & 
- \left( 2\kappa {\dot \Sigma}^{\dagger a} + g f^{abc} \Phi^{\dagger b} \Sigma^{\dagger c} \right)      \ . 
\ena
Under this extended local gauge transformation, the stochastic action (\ref{eq:invariant-stochastic-action-eq2}) is invariant. Once we fix the gauge by specifying the Zwanziger's gauge fixing functions with (\ref{eq:stochastic-gauge-fixing-eq1}), the extended local gauge invariance is broken. In the gauge fixed stochastic action, it is possible to introduce the BRST symmetry.\cite{Nakazawa2} In fact, the stochastic action (\ref{eq:invariant-stochastic-action-eq2}) and the extended local gauge invariance (\ref{eq:extended-local-gauge-transf-eq3}) are the supersymmetric extension of the YM case where SQM is formulated as a 5-dimensional gauge field theory\cite{CH} and the 5-dimensional local gauge invariance leads the BRST symmetry\cite{KOT}\cite{Nakazawa3}. In this note, we mainly consider the stochastic BFM without the BRST symmetry. The perturbative analysis and the renormalization procedure are demonstrated by respecting the background local gauge invariance rather than the BRST symmetry. In \S 7, we extend the BRST invariant formulation of the standard SQM to the stochastic BFM for the proof of the equivalence of the sothcastic BFM to the standard SQM approach.

%\newpage
%%%%%%%%%%%%%%%%%%%%%%%%%%%%%%%%%%%%%%%%%%%%%%%%%%%%%%%%%%%%%%%%%%%%%%%%%%%%%%
%\section{Background field method for (SYM)$_4$ in SQM approach}
%%%%%%%%%%%%%%%%%%%%%%%%%%%%%%%%%%%%%%%%%%%%%%%%%%%%%%%%%%%%%%%%%%%%%%%%%%%%%%

\section{Background field method (BFM) for SYM$_4$ in SQM}

In this section, we formulate the stochastic BFM and apply it to SYM$_4$ following two steps. (I): We consider the background-quantum splitting of the original vector superfield ${\hat V}$ and the auxiliary superfield ${\hat \Pi}$. There exist two types of the local gauge transformations, one is the quantum type and the other is the background one. We fix the quantum type and preserve the background one in the stochastic gauge fixing procedure. (II): We expand the non-polynomial stochastic action with respect to the quantum fluctuations by preserving the background local gauge invariance in each order of the quantum fluctuations. The possible counterterms for the stochastic action are obtained by imposing the backgound local gauge invariance. A consequence of the Ward-Takahashi identities is also obtained in this context. In the next section, we demonstrate the one-loop renormalization of the stochastic action in BFM. We also obtain the one-loop $\beta$-function for the gauge coupling. 

We first splitt the original vector superfield ${\hat V}$ and the original auxiliary superfield ${\hat \Pi}$ in (\ref{eq:invariant-stochastic-action-eq1}) into their quantum flactuations, $V$ and $\varpi$, and their background configurations, $\Omega$, $\Omega^\dagger$ and ${\bf \Pi}$ respectively, 
\bea
\label{eq:background-quantum-splitt-eq1}
e^{2g{\hat V}} 
& \equiv & 
e^{g\Omega}e^{2gV} e^{g\Omega^\dagger}  \ , \nn
{\hat \Pi} 
& \equiv & 
e^{g L_\Omega}{\bf \Pi} + e^{g L_\Omega} \displaystyle{\frac{2g L_V}{1- e^{-2g L_V}}}\varpi   \ . 
\ena
The conventions in the stochastic BFM is summarized in Appendix A. 
The background-quantum splitting for the vector superfield is a conventional one.\cite{GSR} In particular, the background vector superfield {\bf V} is defined by 
$e^{2g{\bf V}} 
 \equiv  
e^{g\Omega} e^{g\Omega^\dagger}$. 
We comment on the background-quantum splitting of the auxiliary superfield, ${\hat \Pi}$, which is essentially a canonical conjugate momentum of the vector superfield ${\hat V}$ with respect to the stochastic time. 

Let us consider a background-quantum splitting ${\hat \Pi} \equiv 
e^{g L_\Omega}{\bf \Pi} + e^{g L_\Omega} {\tilde \varpi} $. 
Under the local gauge transformation, the quantum vector superfield $V$ and 
the background superfield $\Omega$, $\Omega^\dagger$ are transformed as 
\bea
\label{eq:background-local-gauge-transf-eq0}
V 
& \rightarrow &
e^{-ig K}V e^{ig K}      \  , \nn
e^{g \Omega} 
& \rightarrow &
e^{-ig{\hat \Sigma}^\dagger} e^{g \Omega} e^{ig K}   \ , \nn       
e^{g \Omega^\dagger}
& \rightarrow &
e^{- ig K}  e^{g \Omega^\dagger} e^{ig {\hat \Sigma}}      \ .      
\ena
Here the transformation parameter $K$ is a vector superfield $K^\dagger = K$. 
The superfields 
${\hat \Sigma}$ and ${\hat \Sigma}^\dagger$ are chiral and anti-chiral, respectively. (\ref{eq:background-local-gauge-transf-eq0}) is also expressed as 
$
e^{g L_\Omega}
\rightarrow 
e^{-ig L_{{\hat \Sigma}^\dagger}} e^{g L_\Omega} e^{ig L_K}            
$. 
 (See also appendix A.) 
The original auxiliary superfield ${\hat \Pi}$ is transformed as 
$
{\hat \Pi}
\rightarrow 
e^{-ig{\hat \Sigma}^\dagger}{\hat \Pi}e^{ig {\hat \Sigma}^\dagger} 
\equiv  e^{-ig L_{{\hat \Sigma}^\dagger}}{\hat \Pi}              
$
. 
 Therefore, the background field ${\bf \Pi}$ and the quantum fluctuation 
${\tilde \varpi}$ are transformed as 
\bea
\label{eq:background-local-gauge-transf-eq1}
{\bf \Pi}
\rightarrow 
e^{-ig K}{\bf \Pi} e^{ig K} \ , \quad 
{\tilde \varpi}
\rightarrow 
e^{-ig K}{\tilde \varpi}e^{ig K}   \ ,
\ena
under the background local gauge transformation.

Under the quantum local gauge transformation, 
the quantum vector superfield $V$ is transformed by 
$e^{2gV} \rightarrow e^{-ig \Sigma^\dagger} e^{2gV} e^{ig \Sigma}$. 
Here $\Sigma^\dagger \equiv e^{- gL_\Omega}{\hat \Sigma}^\dagger$ ( $\Sigma \equiv e^{gL_{\Omega^\dagger}} {\hat \Sigma}$ ) is a background anti-chiral ( chiral ) superfield  defined by ${\cal D}_\a \Sigma^\dagger = 0$ ( ${\overline {\cal D}}_{\dot \a} \Sigma = 0$ ), where ${\cal D}_\a$ and ${\overline {\cal D}}_{\dot \a}$ are the background covariant spinor derivatives defined by (\ref{eq:background-covariant-derivative-A-eq1}). 
While the background fields $\Omega$  and $\Omega^\dagger$ are invariant, which implies
\bea
\label{eq:quantum-local-gauge-transf-eq1}
{\bf \Pi}
\rightarrow 
e^{-ig L_{\Sigma^\dagger}}{\bf \Pi} \ , \quad 
{\tilde \varpi}
\rightarrow 
e^{-ig L_{\Sigma^\dagger}}{\tilde \varpi}   \ .
\ena
This indicates that the quantum fluctuation ${\tilde \varpi}$ is not a vector superfield. 

Since we prefer to choose the quantum fluctuation as a vector superfield which is the canonical conjugate momentum of $V$, we introduce a vector superfield $\varpi$ by the redefinition, 
${\tilde \varpi} = \displaystyle{\frac{2g L_V}{1- e^{-2g L_V}}}\varpi $, 
as we have done in (\ref{eq:invariant-stochastic-action-eq2}). 
Under the background local gauge transformation, 
it is transformed as 
\bea
\label{eq:background-local-gauge-transf-eq2}
\varpi
\rightarrow 
e^{-ig K}\varpi e^{ig K}     \ .
\ena
On the other hand, under the quantum local gauge transformation, its transformation property is complicated. In the infinitesimal form, the vector superfields $V$ and $\varpi$ are transformed as 
\bea
\label{eq:quantum-local-gauge-transf-eq2}
\delta V^a 
& = & 
+ \displaystyle{\frac{i}{2}} L^a_{\ b} \Sigma^b 
- \displaystyle{\frac{i}{2}} {\Sigma^\dagger}^b L_b^{\ a}   \ , \nn 
\delta \varpi_a 
& = & 
- \displaystyle{\frac{i}{2}} \pa_a L^c_{\ b} \varpi_c \Sigma^b 
+ \displaystyle{\frac{i}{2}} {\Sigma^\dagger}^b \pa_a L_b^{\ c} \varpi_c  
             \ .
\ena
Here the transformation property is derived from the original transformation given in (\ref{eq:extended-local-gauge-transf-eq3}). We notice, in (\ref{eq:extended-local-gauge-transf-eq3}), the transformation is defined for the original superfields, ${\hat V}$ and ${\hat \varpi}$, and the transformation parameters must be understood as ${\hat \Sigma}$ and ${\hat \Sigma}^\dagger$ in the notation of this section. Therefore, the background local gauge transformation is given by simply replacing 
$( {\hat V}, {\hat \varpi}, {\hat \Sigma}, {\hat \Sigma}^\dagger ) \rightarrow ( V, \varpi, \Sigma, \Sigma^\dagger )$ in (\ref{eq:extended-local-gauge-transf-eq3}). 

As explained in Appendix B, the auxiliary superfields for the Zwanziger's gauge fixing functions, ${\hat \Phi}$ and ${\hat {\bar \Phi}}$, are also redefined in such a way that they are background covariantly chiral and background covariantly anti-chiral
\bea
\label{eq:background-invariant-gauge-fixing-eq1}
 \phi \equiv e^{g L_{\Omega^\dagger}} {\hat \Phi}  \ , \  
 {\bar \phi} \equiv e^{- g L_\Omega} {\hat {\bar \Phi}}     \ . 
\ena
${\hat {\bar \Phi}} = {\hat \Phi}^\dagger$ implies 
${\bar \phi} = \phi^\dagger$.
These superfields satisfy the backgound chiral and anti-chiral conditions.
\bea
\label{eq:background-invariant-gauge-fixing-eq2}
{\overline {\cal D}}_{\dot \a} \phi = {\cal D}_\a {\bar \phi} = 0    \ .
\ena
They are simply transformed under the background local gauge transformation 
\bea
\label{eq:background-local-gauge-transf-eq3}
\phi
& \rightarrow &
e^{-ig K}\phi e^{ig K}     \   , \nn
{\bar \phi} 
& \rightarrow &
e^{-ig K}{\bar \phi} e^{ig K}     \ . 
\ena
The transformation property under the quantum local gauge transformation is determined from (\ref{eq:extended-local-gauge-transf-eq3}) by replacing,  
$( {\hat \Phi}, {\hat {\bar \Phi}}, {\hat \Sigma}, {\hat \Sigma}^\dagger ) \rightarrow ( \phi, {\bar \phi}, \Sigma, \Sigma^\dagger )$,  
\bea
\label{eq:quantum-local-gauge-transf-eq3}
\delta \phi^a 
& = & 
- \left( 2\kappa {\overline {\cal D}}_\t \Sigma^a + g f^{abc} \phi^b \Sigma^c 
\right)     \ , \nn        
\delta {\bar \phi}^a 
& = & 
- \left( 2\kappa {\cal D}_\t \Sigma^{\dagger a} + g f^{abc} {\bar \phi}^b \Sigma^{\dagger c} \right)      \ . 
\ena
Here we have introduced the background covariant derivatives with respect to the stochastic time, ${\cal D}_\t$ and ${\overline {\cal D}}_\t$ defined by (\ref{eq:background-covariant-time-derivarive-A-eq1}). 

The quantum local gauge invariance, (\ref{eq:quantum-local-gauge-transf-eq2}) and (\ref{eq:quantum-local-gauge-transf-eq3}), is broken by the following background local gauge invariant stochastic gauge fixing procedure.
We fix the gauge by specifying the Zwanziger's gauge fixing functions, $\phi$ and ${\bar \phi}$. They must be invariant under the background local gauge transformation. The possible extension of the standard Zwanziger's gauge fixing functions (\ref{eq:stochastic-gauge-fixing-eq1}) is almost uniquely determined from the conditions, (\ref{eq:background-invariant-gauge-fixing-eq2}) and 
${\bar \phi} \equiv \phi^\dagger$, and the transformation property (\ref{eq:background-local-gauge-transf-eq3}), as follows.
\bea
\label{eq:background-stochastic-gauge-fixing-eq1}
\phi 
& = & 
 i \displaystyle{\frac{\xi}{4}} {\overline {\cal D}}^2 {\cal D}^2 V     \ , \nn
{\bar \phi} 
& = & 
 - i \displaystyle{\frac{\xi}{4}} {\cal D}^2 {\overline {\cal D}}^2 V  \ . 
\ena
Even after the gauge is fixed, i.e. the Zwanziger's gauge fixing functions are specified by (\ref{eq:background-stochastic-gauge-fixing-eq1}), the stochastic action is background local gauge invariant, provided that the parameters of the background local gauge transformation are restricted to be stochastic time $independent$ 
\bea
\label{eq:background-local-gauge-transf-eq4}
{\dot K} = {\dot {\hat \Sigma}} = {\dot {{\hat \Sigma}^\dagger}} = 0   \ . 
\ena
The condition also implies
${\overline {\cal D}}_\t \Sigma = {\cal D}_\t \Sigma^\dagger = 0$. 

By substituting the definition of the background-quantum splitting (\ref{eq:background-quantum-splitt-eq1}), and using the conventions described in Appendix A,  (\ref{eq:invariant-stochastic-action-eq1}) reads
\bea
\label{eq:invariant-stochastic-action-eq3} 
Z 
& = &  
\int {\cal D}V {\cal D}\varpi e^{K}        \ , \nn
K 
& \equiv & 
\!\!\! \int \!\!\! d^8z d\t {\rm Tr} \Biggl[ 
 \displaystyle{\frac{1}{2\kappa}} 
 \left( {\bf \Pi} + \displaystyle{\frac{2g L_V}{1- e^{-2g L_V}}}\varpi \right)^2           \nn 
 & {} &
+i \left( {\bf \Pi} + \displaystyle{\frac{2g L_V}{1- e^{-2g L_V}}}\varpi \right)\cdot \Biggl( 
\displaystyle{\frac{1}{2g}} 
\Big( \displaystyle{\frac{d}{d\t}} e^{2gV} 
\Big) e^{-2gV}  
+ \displaystyle{\frac{1}{2g}} \left( 
{\cal D}'_\t - e^{2g L_V} {\overline {\cal D}}'_\t 
\right)        \nn
& {} & 
\quad
- \displaystyle{\frac{i}{4\kappa}}   \left( 
{\bar \phi} - e^{2g L_V} \phi \right) 
- \displaystyle{\frac{1}{4\kappa g}}  \left( 
e^{2g L_V} \{ {\nabla}^\a,\ W_\a \} + \{ {\overline \nabla}_{\dot \a},\ {\overline W}^{\dot \a} \} \right) 
\Biggr) 
\Biggr]       \ .
\ena
Here we have defined ${\cal D}_\t' \equiv {\cal D}_\t - \displaystyle{\frac{d}{d\t}}$ and ${\overline {\cal D}}_\t' \equiv {\overline {\cal D}}_\t - \displaystyle{\frac{d}{d\t}}$. 
This is invariant under the stochastic time $independent$ background local gauge transformations 
(\ref{eq:background-local-gauge-transf-eq0}), 
(\ref{eq:background-local-gauge-transf-eq1}), 
(\ref{eq:background-local-gauge-transf-eq2}) and  
(\ref{eq:background-local-gauge-transf-eq3}) 
with the condition, 
(\ref{eq:background-local-gauge-transf-eq4}). 
In the following, we only consider the background local gauge invariance in the restricted sense, i.e. the stochastic time $independent$ one (\ref{eq:background-local-gauge-transf-eq4}). 

For the perturbative analysis, we expand the stochastic action 
(\ref{eq:invariant-stochastic-action-eq3}) 
with respect to the quantum vector superfield $V$ and its canonical conjugate momentum $\varpi$. 
The 0-th order terms with respect to $\varpi$ and $V$ provide a \lq\lq classical \rq\rq stochastic action for background fields. 
\bea
\label{eq:invariant-classical-stochastic-action-eq1} 
K^{(0)} 
& = & 
\int\!\!\!d\t d^8z {\rm Tr} \Big[ 
\displaystyle{\frac{1}{2\kappa}}{\bf \Pi}^2 
+ \displaystyle{\frac{i}{2g}} {\bf \Pi} 
\Bigl(
( \displaystyle\frac{d}{d\t}e^{g\Omega^\dagger} )e^{-g\Omega^\dagger} 
+ e^{-g\Omega} ( \displaystyle\frac{d}{d\t}e^{g\Omega} ) 
\Bigr)          \nn
& {} & \qquad\qquad\qquad\qquad\qquad 
- \frac{i}{4\kappa g} {\bf \Pi} \Bigl( \{ {\cal D}^\a,\ W_\a^{(0)} \} 
+  \{ {\overline {\cal D}}_{\dot \a},\ {\overline W}^{{\dot \a} (0)} \}  
\Bigr)
\Big]      \ .
\ena
The classical stochastic action is, of course, invariant under the background local gauge transformation. As we discuss later, this classical action defined by the background fields also determines possible counterterms for the renormalization procedure. 

The first order terms with respect to $\varpi$ and $V$ may provide the field equations for the background fields  $\Omega$, $\Omega^\dagger$ and ${\bf \Pi}$ in a conventional sense. 
\bea
\label{eq:background-classical-equation-eq1} 
\displaystyle{\frac{1}{\kappa}} {\bf \Pi} 
& + & 
 \displaystyle{\frac{i}{2g}}  \Big(
( \displaystyle\frac{d}{d\t}e^{g\Omega^\dagger} )e^{-g\Omega^\dagger} 
+ e^{-g\Omega} ( \displaystyle\frac{d}{d\t}e^{g\Omega} ) 
\Big)                                \nn
& {} & \qquad\qquad\qquad 
- \frac{i}{4\kappa g}  \Big( 
\{ {\cal D}^\a,\ W_\a^{(0)} \} 
+  \{ {\overline {\cal D}}_{\dot \a},\ {\overline W}^{{\dot \a} (0)} \}  
\Big)    = 0 \ , \nn
\displaystyle\frac{d}{d\t} {\bf \Pi} 
& + & 
[\ {\bf \Pi},\  \displaystyle{\frac{d}{d\t}}( e^{g\Omega^\dagger} )e^{-g\Omega^\dagger}\  ] 
+ \displaystyle{\frac{\xi}{16\kappa}} 
\left( {\cal D}^2{\overline {\cal D}}^2 + {\overline {\cal D}}^2 {\cal D}^2 \right) {\bf \Pi} \nn
& {} & \qquad\qquad\qquad
- \displaystyle{\frac{1}{\kappa}} {\overline {\cal D}}_{\dot \a} 
[\ {\bf \Pi},\ {\overline W}^{{\dot \a} (0)}\ ] 
- \displaystyle{\frac{1}{8\kappa}} {\overline {\cal D}}_{\dot \a}{\cal D}^2 {\overline {\cal D}}^{\dot \a} {\bf \Pi} = 0 \ .
\ena
In the conventional BFM, in particular to discuss the S-matrix in this context, equations of motion are assumed for the background fields\cite{'tHooft}. In the stochastic BFM, however, it is difficult to extract the S-matirx from the effective stochastic action which is a generating functional of the 1-P-I vertices of the connected stochastic Feynman diagrams. Therefore, in a precise sense, we do $not$ assume the field equations (\ref{eq:background-classical-equation-eq1}) for the background superfields in order to define the generator of the 1-P-I vertices from the connected stochastic Green's functions in a standard manner. In \S 6, we define a reduced form of the effective stochastic action 
${\tilde \Gamma}( {\tilde V}, {\tilde \varpi}, {\bf V}, {\bf \Pi} )$ in the stochastic BFM  by taking the vanishing limit of the expectation values of the quantum superfields, ${\tilde V}= {\tilde \varpi}= 0$. It is shown that the effective stochastic action in this limit, ${\tilde \Gamma}( 0, 0, {\bf V}, {\bf \Pi} )$, is equivalent to the standard effective stochastic action defined in an unusual stochastic gauge fixing.  
 
The second order terms with respect to $\varpi$ and $V$ provide the kinetic term which defines the superpropagators 
\bea
\label{eq:stochastic-action-kinetic-eq1} 
K^{(2)}_{\rm free} 
& = & \int\!\!\!d\t d^8z {\rm Tr} \Big[ 
\displaystyle{\frac{1}{2\kappa}} \varpi^2   
+ i\varpi \Big\{ 
{\dot V} 
+ \displaystyle{\frac{1}{8\kappa}}{\overline D}_{\dot \a} D^2 {\overline D}^{\dot \a}  V           \nn 
& {} & \qquad\qquad\qquad\qquad\qquad \qquad 
- \displaystyle{\frac{\xi}{16\kappa}} 
\left( D^2 {\overline D}^2 + {\overline D}^2 D^2 \right) V 
\Big\} 
\Big]      \  , 
\ena
and the interaction terms relevant to the one-loop perturbation, 
\bea
\label{eq:stochastic-action-interaction-eq1} 
K^{(2)}_{\rm int} 
& = & \int\!\!\!d\t d^8z {\rm Tr} \Big[ 
\displaystyle{\frac{g}{\kappa}} {\bf \Pi} [\ V,\ \varpi\ ] 
+ ig {\bf \Pi} 
\Big\{  
[\ V,\ {\dot V}\ ] - [\ V,\ [\ V,\ {\overline {\cal D}}'_\t\ ]\ ] 
+ \displaystyle{\frac{i}{2\kappa}} [\ V,\ \phi\ ]      \nn
& {} & 
+ \displaystyle{\frac{1}{\kappa}}\{\ 
[\ V,\ {\overline {\cal D}}_{\dot \a}V\ ],\ {\overline W}^{{\dot \a} (0)}\ 
 \}
- \displaystyle{\frac{1}{4\kappa}}\{\ 
 {\overline {\cal D}}_{\dot \a} V,\ 
 {\cal D}^2 {\overline {\cal D}}^{\dot \a} V\ 
 \} 
+ \displaystyle{\frac{1}{8\kappa}} 
 {\overline {\cal D}}_{\dot \a}{\cal D}^2 [\ V,\ {\overline {\cal D}}^{\dot \a} V\ ] 
\Big\}    \nn
& {} & 
+ i \varpi 
\Big\{ 
- \displaystyle{\frac{1}{2}}[\ V,\ {\cal D}'_\t + {\overline {\cal D}}'_\t\ ]
- \displaystyle{\frac{i}{4\kappa}} ( {\bar \phi}' - \phi' ) 
+ \displaystyle{\frac{1}{\kappa}} \{\ 
{\overline {\cal D}}^{\dot \a} V,\  {\overline W}^{{\dot \a} (0)}\ 
\}                \nn
& {} & 
+ \displaystyle{\frac{1}{8\kappa}}  \left( {\overline {\cal D}}_{\dot \a} {\cal D}^2 {\overline {\cal D}}^{\dot \a} 
- {\overline D}_{\dot \a} D^2 {\overline D}^{\dot \a} \right) V   
+ \displaystyle{\frac{1}{2\kappa}} [\ 
V,\ {\overline {\cal D}}_{\dot \a} {\overline {\cal W}}^{{\dot \a} (0)}\ ] 
\Big\}
\Big]         \ . 
\ena
Here 
we have defined 
$\phi'\equiv \phi -  i \displaystyle{\frac{\xi}{4}} {\overline D}^2 D^2 V  $ 
and 
${\bar \phi}'
 \equiv {\bar \phi} + i \displaystyle{\frac{\xi}{4}} D^2 {\overline D}^2 V  $. 

The kinetic term $K^{(2)}_{\rm free}$ defines the superpropagators 
\bea
\label{eq:stochastic-action-kinetic-eq2} 
K^{(2)}_{\rm free} 
& = & \int\!\!\!d\t d^8z {\rm Tr} \Big[  
\displaystyle{\frac{1}{2\kappa}} \varpi^2   
+ i\varpi \Big\{ 
{\dot V} 
- \displaystyle{\frac{1}{\kappa}} \square V 
+ \displaystyle{\frac{1 - \xi}{16\kappa}} 
\left( D^2 {\overline D}^2 + {\overline D}^2 D^2 \right) V 
\Big\} 
\Big]     \ .
\ena
In the Feynman gauge $\xi = 1$, we obtain the simplest forms of the superpropagators as follows. 
\bea
\label{eq:super-propagator-eq1} 
\langle \varpi^a ( \t, z ) \varpi^b ( \t', z' ) \rangle 
& = & 0              \ , \nn 
\langle V^a ( \t, z ) \varpi^b ( \t', z' ) \rangle  
& = & 
\delta^{ab}  
\displaystyle{\frac{i}{(  i\omega + k^2/\kappa )}} 
\delta^2 ( \theta - \theta' ) \delta^2 ( {\bar \theta} - {\bar \theta'} ) 
          \ , \nn  
\langle V^a ( \t, z ) V^b ( \t', z' ) \rangle  
& = &
-  \displaystyle{\frac{\delta^{ab}}{\kappa}}
\displaystyle{\frac{1}{( i\omega + k^2/\kappa )( - i\omega + k^2/\kappa )}}  
\delta^2 ( \theta - \theta' ) \delta^2 ( {\bar \theta} - {\bar \theta'} ) 
       . 
\ena
Here we have suppressed the momentum integration, 
$
\int \!\!\! d^4k d\omega (2\varpi)^{-5} 
e^{ik \cdot ( x - x' ) + i \omega \cdot ( \t - \t' )}
$ .
For the one-loop renormalization in the stochastic BFM, we determin the possible counterterms by the background local gauge invariance and the dimensional analysis. We note
\bea
\label{eq:dimensional-analysis-eq1}
 [ x^m ] 
& = & -1,\ [ \theta^\a ] = [ {\bar \theta}^{\dot \a}]  = - \displaystyle{\frac{1}{2}},\  [ \t ] = -2, \ 
 [ {\cal D}_\a ] = [ {\overline {\cal D}}_{\dot \a} ] = \displaystyle{\frac{1}{2}}    \ ,    [ {\cal D}_\t ] = [ {\overline {\cal D}}_\t ] = 2 \ ,   \nn     
 {[} \Omega ] 
& = & [ \Omega^\dagger ] = [ {\bf V} ] = 0,\ 
[ {\bf \Pi} ] = 2,\ 
[ W_\a^{(0)} ] = [ {\overline W}_{\dot \a}^{(0)} ] = \displaystyle{\frac{3}{2}}      \  . 
\ena
The derivative with respect to the stochastic time must appear as the covariant derivatives ${\cal D}_\t$ and ${\overline {\cal D}}_\t$. 
If we assume in general the possible counterterms as 
\bea
\label{eq:dimensional-analysis-eq2} 
( {\bf  \Pi} )^p ( {\cal D}_\t )^q ( {\overline {\cal D}}_\t )^r ( {\cal D}_\a )^s 
( {\overline {\cal D}}_{\dot \a} )^x ( W_\a^{(0)} )^y ( {\overline W}_{\dot \a}^{(0)})^z     \ , 
\ena
we obtain a condition 
\bea
\label{eq:dimensional-analysis-eq3} 
2( p + q + r ) + \displaystyle{\frac{1}{2}} ( s + x ) + \displaystyle{\frac{3}{2}} ( y + z ) = 4       \ .
\ena
From the retarded nature of the superpropagator, $\langle V \varpi \rangle$, the relevant stochastic Feynman diagrams include at least one background auxiliary superfield, ${\bf \Pi}$, as the external line. Therefore we have only three cases.
\bea
\label{eq:dimensional-analysis-eq4} 
p & = & 2 \ , \  {\bf \Pi}^2  \ , \nn
p & = & 1 \ , \  q = 1 \ ( {\rm or}\ r =1 )  \ , \       
{\cal D}_\t {\bf \Pi} \ , \ {\overline {\cal D}}_\t  {\bf \Pi} \ , \ \nn   
p & = & 1 \ , \  s = y = 1 \ ( {\rm or}\ x = z = 1) \ , \  
{\bf \Pi} {\cal D}^\a  W_\a^{(0)}  \ , \ {\bf \Pi} {\overline {\cal D}}_{\dot \a} {\overline W}^{{\dot \a} (0)}       \ .   
\ena
The reality condition indicates that 
$ {\cal D}^\a  W_\a^{(0)}  = {\overline {\cal D}}_{\dot \a} {\overline W}^{{\dot \a} (0)}$. In the case, $p  =  1 \ , \  s = 2 \ , \ x = 2$, we would have 
\bea
\label{eq:dimensional-analysis-eq5} 
& {} & 
{\cal D}^\a {\overline {\cal D}}^2 {\cal D} {\bf \Pi} \ , \ 
{\overline {\cal D}}_{\dot \a} {\cal D}^2 {\overline {\cal D}}^{\dot \a} {\bf \Pi} \ , \ 
{\cal D}^\a {\overline {\cal D}}_{\dot a} {\cal D}_\a {\overline {\cal D}}^{\dot \a} {\bf \Pi} \ , \nn
& {} &
{\overline {\cal D}}_{\dot \a}{\cal D}^\a {\overline {\cal D}}^{\dot \a} {\cal D}_\a {\bf \Pi}              \ , \ 
{\cal D}^2 {\overline {\cal D}}^2 {\bf \Pi} \ , \ 
{\overline {\cal D}}^2 {\cal D}^2 {\bf \Pi} \ .
\ena
However, the covariant spinor derivatives, ${\cal D}_\a$ and ${\overline {\cal D}}_{\dot \a}$, must appear as ${\cal D}^\a W_\a$, ${\overline {\cal D}}_{\dot \a} {\overline W}^{\dot \a}$, $W^\a {\cal D}_\a {\bf \Pi}$, ${\overline W}_{\dot \a} {\overline {\cal D}}^{\dot \a} {\bf \Pi}$ and its anti-commutator, ${\cal D}_m$, which leads ${\cal D}^m{\cal D}_m {\bf \Pi}$ in the perturbative calculation with the covariantized $D$-algebra\cite{GZ}. Since ${\cal D}^m{\cal D}_m {\bf \Pi}$ must not appear because its leading term is $\square {\bf \Pi}$, the remaining terms are reduced to (\ref{eq:dimensional-analysis-eq4}) by partial integrations. 
  The covariant derivative term with respect to the stochastic time must be 
$
( {\cal D}_\t - {\overline {\cal D}}_\t  ){\bf \Pi}  \ ,
$
because  
$
\displaystyle{\frac{d}{d\t}} {\bf \Pi} 
$
 does not appear. 
Finaly we arrive at the following counterterms.
\bea
\label{eq:counter-term-eq1} 
K^{(0)} 
& = & 
\int\!\!\!d\t d^8z {\rm Tr} \Biggl( 
c_1 {\bf \Pi}^2 
+ i c_2  ( {\cal D}_\t - {\overline {\cal D}}_\t ) {\bf \Pi}          \nn
& {} & \qquad\qquad\qquad 
+i c_3 {\bf \Pi} \biggl( \{ {\cal D}^\a,\ W_\a^{(0)} \} 
+  \{ {\overline {\cal D}}_{\dot \a},\ {\overline W}^{{\dot \a} (0)} \} 
\biggr) 
\Biggr)      \ .
\ena
We show by the power counting argument that these counterterms are sufficient to cancel the ultra-violet divergences in all order of the perturbation in the last of this section. 
(\ref{eq:counter-term-eq1}) is nothing but the classical stochastic action for the background fields (\ref{eq:invariant-classical-stochastic-action-eq1}). By multiplicative renormalizations 
\bea
\label{eq:renormalization-constant-eq1} 
{\bf V}_{\rm bare} = \sqrt{Z_V} {\bf V} \ , \ 
{\bf \Pi}_{\rm bare} = \sqrt{Z_\Pi} {\bf \Pi} \ , \ 
g_{\rm bare} = Z_g g \ , \ 
\kappa_{\rm bare} = Z_\kappa \kappa  \ , 
\ena
the classical action for the background fields is redefined as
\bea
\label{eq:counter-term-eq2} 
K^{(0)} 
& = & 
\int\!\!\!d\t d^8z {\rm Tr} \Biggl( 
Z_\kappa^{-1} Z_\Pi \displaystyle{\frac{1}{2\kappa}}{\bf \Pi}^2 
+ 
Z_g^{-1} \sqrt{Z_\Pi} \displaystyle{\frac{i}{2g}} {\bf \Pi} \biggl( 
( \displaystyle\frac{d}{d\t}e^{g\Omega^\dagger} )e^{-g\Omega^\dagger} 
+ e^{-g\Omega} ( \displaystyle\frac{d}{d\t}e^{g\Omega} ) 
\biggr)          \nn
& {} & \qquad\qquad\qquad 
- 
Z_\kappa^{-1} Z_g^{-1} \sqrt{Z_\varpi} \frac{i}{4\kappa g} {\bf \Pi} \biggl( \{ {\cal D}^\a,\ W_\a^{(0)} \} 
+  \{ {\overline {\cal D}}_{\dot \a},\ {\overline W}^{{\dot \a} (0)} \}  
\biggr) 
\Biggr)      \ .
\ena
Here we have used the following Ward-Takahashi identity
\bea
\label{eq:Ward-Takahashi-eq1} 
Z_g \sqrt{Z_V} = 1  \ , 
\ena
which ensures $\Omega_{\rm bare} = \sqrt{Z_{V}} \Omega$ and $\Omega^\dagger_{\rm bare} = \sqrt{Z_{V}} \Omega^\dagger$. 
This is a consequence of the background local gauge invariance of the stochastic action (\ref{eq:counter-term-eq2}) and a well-known relation in the path-integral approach. 
By expanding this classical action with respect to the background fields in the gauge $\Omega = \Omega^\dagger = {\bf V}$, we obtain 
\bea
\label{eq:counter-term-eq3} 
K^{(0)} 
& = & 
\int\!\!\!d\t d^8z {\rm Tr} \Biggl( 
Z_\kappa^{-1} Z_\Pi \displaystyle{\frac{1}{2\kappa}}{\bf \Pi}^2 
+ 
i \sqrt{Z_\Pi Z_V} {\bf \Pi} {\dot {\bf V}}     \nn
& {} & \qquad\qquad\qquad\qquad 
+ 
Z_\kappa^{-1} \sqrt{Z_\Pi Z_V} \frac{i}{8\kappa} 
{\bf \Pi} D^\a {\overline D}^2 D_\a {\bf V}   + ... 
\Biggr)      \ .
\ena
The renormalization of the gauge parameter $\xi$ is not necessary for the one-loop perturbation. For higher loop calculation, we introduce the renormalization constant for the gauge parameter $\xi_{\rm bare} = Z_\xi \xi$.

%\newpage
%%%%%%%%%%%%%%%%%%%%%%%%%%%%%%%%%%%%%%%%%%%%%%%%%%%%%%%%%%%%%%%%%%%%%%%%%%%%%%
%\section{Perturbations in the BFM for (SYM)$_4$ 
%in SQM}
%%%%%%%%%%%%%%%%%%%%%%%%%%%%%%%%%%%%%%%%%%%%%%%%%%%%%%%%%%%%%%%%%%%%%%%%%%%%%%

\section{One-loop perturbation in the stochastic BFM} 

The perturbative calculation is drastically simplified by the background local gauge invariance. 
We employ the regularization via the dimensional reduction\cite{Siegel}\cite{CJN} which preserves the global supersymmetry as well as the background local gauge symmetry in the superfield formalism. In the dimensional reduction, the so-called $D$-algebra is carried out in the four dimensional sense. After reducing the covariant spinor derivatives into their anti-commutator except four spinor derivatives, $D_\a D_\b {\overline D_{\dot \a}}{\overline D_{\dot \b}}$, which are necessary to eliminate $\delta^2( \theta )\delta^2( {\bar \theta} )$, the momentum integration is performed by the analytic continuation to $d$ dimensions. We here use the standard $D$-algebra\cite{GSR} in the explicit calculation instead of the covariantized one\cite{GZ}. There must be at least four covariant spinor derivatives, $D_\a D_\b {\overline D_{\dot \a}}{\overline D_{\dot \b}}$, for non-vanishing contribution. 
At first we note that, in the one-loop calculation, the contribution of $\langle K^{(2)}_{\rm int} \rangle$ is finite in the regularization via the dimensional reduction. 
The possible contribution to divergences begins from $\langle K^{(2)}_{\rm int} K^{(2)}_{\rm int} \rangle$. In the stochastic BFM, by assuming the background local gauge invariance, it is not necessary to evaluate all the possible diagrams. In order to simplify the calculation, we only pick up the lowest order in the expansion with respect to the background fields. This means that the stochastic Feynman diagrams inlcude at most two background fields as the external lines. 

By the expansion with respect to the backgound fields, the interaction part of the stochastic action (\ref{eq:stochastic-action-interaction-eq1}) is reduced to 
\bea
\label{eq:one-loop-interaction-eq1}
K^{(2)} 
& = & 
\int \!\!\! d\t d^8z 
\Biggl(  
i\displaystyle{\frac{g}{\kappa}} {\bf \Pi}\cdot V \times \varpi  
+ i\displaystyle{\frac{g}{4\kappa}} {\bar \sigma}^{m {\dot \beta}\a} 
\varpi \cdot ( \pa_m {[}\ D_\a,\ {\overline D}_{\dot \beta}\ ]{\bf V} ) \times V \nn
& {} & 
+ i\displaystyle{\frac{g}{2\kappa}} {\bar \sigma}^{m {\dot \beta}\a} 
\varpi \cdot ( {[}\ D_\a,\ {\overline D}_{\dot \beta}\ ]{\bf V} ) \times \pa_m V 
- \displaystyle{\frac{1}{\kappa}} \varpi \cdot W^{\a (0)} \times D_\a V 
+ \displaystyle{\frac{1}{\kappa}} \varpi \cdot {\overline W}_{\dot \a}^{(0)} \times {\overline D}^{\dot \a} V      \nn
& {} & 
- g {\bf \Pi} \cdot V\times {\dot V}
+ \displaystyle{\frac{g}{8\kappa}}{\bf \Pi} \cdot V\times {\overline D}^2 D^2 V 
+ \displaystyle{\frac{g}{4\kappa}}{\bf \Pi} \cdot {\overline D}_{\dot \a}V\times D^2 {\overline D}^{\dot \a} V     \nn 
& {} &
- \displaystyle{\frac{g}{8\kappa}}{\bf \Pi} \cdot {\overline D}_{\dot \a} D^2 ( V \times {\overline D}^{\dot \a} V )
 \Biggr)       \ .
\ena
Here $A\cdot B\times C \equiv f^{abc} A^a B^b C^c$. 

Amang the possible stochastic Feynman diagrams, 12 diagrams are non-trivial. In these non-trivial diagrams, 8 diagrams are found to be finite. Only the remaining 4 diagrams contribute to the ultra-violet divergences. In the dimensional reduction, the regularization of the momentum integration is the same as the conventional dimensional regularization. The typical regularized integrals are evaluated as 
\bea
\label{eq:one-loop-integral-eq1}
& {} &
\displaystyle{\frac{1}{\kappa}}
\int \!\!\! \displaystyle{\frac{d^dk d\omega}{(2\pi)^5}} 
\displaystyle{\frac{1}{( i\omega + k^2/\kappa )( - i\omega + k^2/\kappa )}} \nn
& {} & \qquad
\times \displaystyle{\frac{1}{( i( \omega - \lambda )  + ( k -q )^2/\kappa )( - i ( \omega -\lambda ) + ( k - q )^2/\kappa )}} k^\mu k^\nu      
= \displaystyle{\frac{\delta^{\mu\nu}}{4}}I_\epsilon      \ , \nn 
& {} & 
\int \!\!\! \displaystyle{\frac{d^dk d\omega}{(2\pi)^5}} 
\displaystyle{\frac{-1}{( i\omega + k^2/\kappa )( - i\omega + k^2/\kappa )}} 
\displaystyle{\frac{i}{(  -i ( \omega - \lambda ) + ( k - q )^2/\kappa )}} 
= - i I_\epsilon      \nn
& {} & 
I_\epsilon 
\equiv 
\displaystyle{\frac{\kappa}{4}}( \displaystyle{\frac{1}{4\pi}} )^2 \Gamma (\epsilon)   \ , \
\ena
where $\epsilon \equiv 2 - \displaystyle{\frac{d}{2}}$. We note, in the regularization via the dimensional reduction, the renormalization constants (\ref{eq:renormalization-constant-eq1}) are redefined as 
$
g_{\rm bare} = \mu^{2\epsilon} Z_g g \ 
$
and 
$
\kappa_{\rm bare} = Z_\kappa \kappa 
$. Namely, $\kappa_{\rm bare}$ is dimensionless. 

The result for the one-loop divergences is given by 
\bea
\label{eq:one-loop-divergence-eq1}
& {} &
\Gamma^{\rm div} ( {\bf V},\ {\bf \Pi} )     \nn
& {} & \quad 
= 
\langle  
K^{(2)}_{\rm int} 
\rangle + 
\displaystyle{\frac{1}{2}} \Big( 
\langle   K^{(2)}_{\rm int} K^{(2)}_{\rm int}  \rangle 
  - 
\langle  
K^{(2)}_{\rm int} 
\rangle^2 
\Big)                            \nn
& {} & \quad 
= 
\int\!\!\!d\t d^8z {\rm Tr} \Biggl( 
 2 C_2 (G) I_\epsilon \displaystyle{\frac{g^2}{\kappa^2}}{\bf \Pi}^2 
+ 
4i C_2 (G) I_\epsilon \displaystyle{\frac{g}{\kappa^2}}{\bf \Pi} 
\{ {\overline D}_{\dot \a},\ {\overline W}^{{\dot \a} (0)} \}   
\Biggr)           \ . 
\ena
The expression may be covariantized to be invariant under the background local gauge transformation
\bea
\label{eq:one-loop-divergence-eq2}
& {} &
\Gamma^{\rm div} ( {\bf V},\ {\bf \Pi} )     \nn
& {} & \quad 
= 
\int\!\!\!d\t d^8z {\rm Tr} \Biggl( 
 2 C_2 (G) I_\epsilon \displaystyle{\frac{g^2}{\kappa^2}}{\bf \Pi}^2 
+ 
4i C_2 (G) I_\epsilon \displaystyle{\frac{g}{\kappa^2}}{\bf \Pi} 
\{ {\overline {\cal D}}_{\dot \a},\ {\overline W}^{{\dot \a} (0)} \}   
\Biggr)          \ . 
\ena

A remarkable observation in the one-loop divergence (\ref{eq:one-loop-divergence-eq2}) is that the derivative term with respect to the stochastic time, ${\bf \Pi} {\dot {\bf V}}$, is not renormalized in the one-loop level. 
By comparing this result to the counterterms in (\ref{eq:counter-term-eq2}),  
the renormalization constants satisfy 
\bea
\label{eq:one-loop-renormalization-eq1}
\sqrt{Z_V Z_\Pi} -1 
& = & 0        \ , \nn 
Z_\kappa^{-1} Z_\Pi - 1 
& = & -4 C_2 (G) I_\epsilon \displaystyle{\frac{g^2}{\kappa}}                        \ , \nn 
Z_\kappa^{-1} \sqrt{Z_V Z_\Pi} - 1 
& = & 8 C_2 (G) I_\epsilon \displaystyle{\frac{g^2}{\kappa}}        \ .
\ena
This reads
\bea
\label{eq:one-loop-renormalization-eq2}
Z_V 
& = & 
1 + 3  g^2  \displaystyle{\frac{C_2 (G)}{(4\pi)^2}} \displaystyle{\frac{1}{\epsilon}}        \ ,   \nn 
Z_\Pi 
& = & 
1 - 3  g^2  \displaystyle{\frac{C_2 (G)}{(4\pi)^2}} \displaystyle{\frac{1}{\epsilon}}         \ ,   \nn
Z_g 
& = & 
1 - \displaystyle{\frac{3}{2}}  g^2  \displaystyle{\frac{C_2 (G)}{(4\pi)^2}}  \displaystyle{\frac{1}{\epsilon}}         \  , \nn 
Z_\kappa
& = & 
1 - 2  g^2  \displaystyle{\frac{C_2 (G)}{(4\pi)^2}}  \displaystyle{\frac{1}{\epsilon}}         \  . 
\ena
The result for the wave function and the gauge coupling constant coincides with that given in the conventional BFM. This leads the well-known one-loop $\beta$-function for $N=1$ SYM$_4$ without chiral matter multiplets.\cite{FZ} 
\bea
\label{eq:one-loop-beta-function-eq1}
\beta ( g ) = - 3 g^3 \displaystyle{\frac{C_2 (G)}{(4\pi)^2}}   \ .
\ena

In addition to this gauge coupling $\beta$-function, in SQM, we obtain the other $\beta$-function for the scaling parameter of the stochastic time, 
$\kappa$ 
\bea
\label{eq:one-loop-beta-function-eq2}
\beta_\kappa ( g ) = - 4 g^2 \displaystyle{\frac{C_2 (G)}{(4\pi)^2}} \ .
\ena
We also list the anomalous dimensions of the wave function renormalizations.
\bea
\label{eq:one-loop-beta-function-eq3}
\gamma_V ( g )
& = & - 3 g^2 \displaystyle{\frac{C_2 (G)}{(4\pi)^2}}      \ , \nn 
\gamma_\Pi ( g )
& = & + 3 g^2 \displaystyle{\frac{C_2 (G)}{(4\pi)^2}}    \ .
\ena

In the conventional BFM, the Faddeev-Popov prescription requires a Nielsen-Kallosh type ghost in addition to the Faddeev-Popov ghost. Furthermore, due to the non-renormalization theorem, the vector loop does not contribute to the self-energy nor to the three-point function of the vector mutiplet in the one-loop level. Therefore the contribution to the self-energy of the vector multiplet only comes from the Fddeev-Popov and the Nielsen-Kallosh ghosts. In this respect, the stochastic gauge fixing procedure in the stochastic BFM simulates the contribution from both the Faddeev-Popov and the Nielsen-Kallosh ghosts in the one-loop level. On the other hand, in the standard SQM approach, the stochastic gauge fixing procedure for SYM$_4$ reproduces the Faddeev-Popov probability distribution given in the standard path-integral of SYM$_4$ for which the Nielsen-Kallosh ghost is not necessary\cite{Nakazawa2}. The stochastic gauge fixing procedure is introduced by adding the generator of the local gauge transformation in the time evolution equation of observables in such a way that the time evolution of the local gauge invariant quantities does not depend on the Zwanziger's gauge fixing functions. Therefore we expect that the background local gauge invariant choice of the Zwanziger's gauge fixing functions reproduces the probability distribution given in the conventional BFM which requires the Nielsen-Kallosh ghost. The one-loop result supports this conjecture. To confirm it more precisely, we give a proof in the next section on the equivalence of the stochastic BFM to the standard SQM in terms of the effective stochastic action which is a generator of the 1-P-I vertices. In \S 7, we also discuss this equivalence in terms of the BRST invariant formulation of SQM. Since the standard SQM is equivalent to the Faddeev-Popov prescription in the standard path-integral approach, the proof may explain the reason why the stochastic BFM reproduces the contribution of the Nielsen-Kallosh ghost as well as the Faddeev-Popov ghost in the same reliablility as that the Faddeev-Popov prescription in the conventional BFM is equivalent to the standard Faddeev-Popov prescription. 

In the last of this section, we comment on the renormalizability of SYM$_4$ in the SQM approach. The power counting argument is essentially the same as given in Ref.\cite{Nakazawa2}. 
The estimation of the divergence of a stochastic Feynman diagram $G( V,I,E )$, which consists of $V$ vertices, $I$ internal lines, $E$ external lines and $L \equiv I - V +1$ loops is as follows. Precisely, the stochastic Feynman diagram is specified by the following quantities, 
\bea
\label{eq:power-counting-eq1}
I \equiv I_{V \varpi} + I_{V V}, \qquad 
E &\equiv& E_{\bf \Pi} + E_{\bf V}.  \nn
E_{\bf \Pi} : \#\ {\rm external\ lines\ of}\ {\bf \Pi}, \quad 
E_{\bf V} &:& \#\ {\rm external\ lines\ of}\ {\bf V}             \nn
I_{V \varpi} : \#\ {\rm internal\ lines\ of}\ \langle V \varpi \rangle,  \quad 
I_{V V} &:& \#\ {\rm internal\ lines\ of}\ \langle V V \rangle     \nn 
V_{1,n}  :  \#\ {\rm vertices\ with}\ \varpi ( {\rm or}\ {\bf \Pi} )\ {\rm and}\ 
& n & V ( {\rm or}\ {\bf V} ), \nn 
V_{2,n} : \#\ {\rm vertices\ with\ two}\ \varpi ( {\rm or}\ {\bf \Pi} )\ {\rm and}\ 
& n & V ( {\rm or}\ {\bf V} ). 
\ena
We have also topological relations
\bea
\label{eq:power-counting-eq2}
\sum_{n=2} n ( V_{1,n} + V_{2,n} ) 
 =  
2 I_{V V} + I_{V \varpi} + E_{\bf V},      \qquad
\sum_{n=2} ( V_{1,n} + 2 V_{2,n} ) 
 =  
I_{V \varpi} + E_{\bf \Pi}          \  .  
\ena
The degree of the ultra-violet divergence of the stochastic Feynman diagram $G( V,I,E )$ is given by 
\bea
\label{eq:power-counting-eq3}
4L -2 I_{V V} -2 I + 2( \sum_{n=2} V_{1,n} )       
 =  4 - 2E_{\bf \Pi}     \ .
\ena 

From the degree of the divergence, the following types of counterterms, which must be invariant under the backgound local gauge transformation, are necessary to cancel the divergences. \\
$E_{\bf \Pi} = 2$, logarithmic divergences 
$\rightarrow$  two ${\bf \Pi}$ and infinite number of 
${\bf V}$;\\
\noindent 
$\qquad  {\bf \Pi}^2 \ .$\\
$E_{\bf \Pi} = 1$, logarithmic divergences 
$\rightarrow$ one ${\bf \Pi}$ and infinite number of ${\bf V}$;\\
$\qquad ( {\cal D}_\t - {\overline {\cal D}}_\t )  {\bf \Pi} $, $ \ {\bf \Pi} \{ {\cal D}^\a,\ [ {\overline {\cal D}}_{\dot \a},\ \{ {\overline {\cal D}}^{\dot \a},\ {\cal D}_\a \} ] \} $, ... \\
$E_{\bf \Pi} = 0$, no relevant stochastic Feynman diagrams.  

\noindent
This power counting argument is consistent to the previous argument based on the dimensional analysis. The symmetry requirements given in the previous section is sufficient to specify the counterterms which cancel these divergences. In particular, for the $E_{\bf \Pi} = 2$ case, the backgound vector superfield ${\bf V}$ ( or $\Omega$ and $\Omega^\dagger$ ) must appear as the covariant derivatives ${\cal D}_\t$, ${\overline {\cal D}}_\t$, ${\cal D}_\a$ and ${\overline {\cal D}}_{\dot \a}$ in the stochastic BFM. Therefore the possible counterterm for $E_{\bf \Pi} = 2$ is only ${\rm Tr} {\bf \Pi}^2$ on dimensional ground. The divergence for the $E_{\bf \Pi} = 1$ case is also reduced to the logarithmic one due to the background local gauge invariance. Hence we conclude that SYM$_4$ in the superfield formalism is renormalizable in the stochastic BFM by means of the stochastic action principle.

%\newpage
%%%%%%%%%%%%%%%%%%%%%%%%%%%%%%%%%%%%%%%%%%%%%%%%%%%%%%%%%%%%%%%%%%%%%%%%%%%%%%
%\section{The equivalence of the BFM to a standard 
%stochastic action principle} 
%%%%%%%%%%%%%%%%%%%%%%%%%%%%%%%%%%%%%%%%%%%%%%%%%%%%%%%%%%%%%%%%%%%%%%%%%%%%%%

\section{The equivalence of the stochastic BFM to the standard SQM for SYM$_4$} 

In this section, we give a proof on the equivalence of the stochastic BFM for SYM$_4$ to the standard SQM. The argument here is based on that given in the conventional BFM\cite{Abbott}\cite{GGRS}. The relation of the stochastic BFM to the standard SQM is clarified in the Yang-Mills case\cite{Okano}\cite{MOSZ}. A slightly improved point of our stochastic BFM formulation in comparison with the Yang-Mills case\cite{Okano}\cite{MOSZ} is that we have introduced the backgound superfield ${\bf \Pi}$ for the auxiliary superfield ${\hat \varpi}$ as well as for the vector superfield ${\hat V}$. As we have shown in the previous two sections, it is necessary to obtain the manifestly background local gauge invariant counterterms. In terms of 
${\hat U} = e^{2g {\hat V}}$, the structure of SYM$_4$ is similar to that of non-linear $\sigma$-models. In applying the stochastic BFM to non-linear $\sigma$-models, we also need a background field for the auxiliary field, a canonical conjugate momentum of the dynamical field, to generate general coordinate invariant counterterms for the renormalization of the stochastic action\cite{NE}. 

 Let us define a generating functional of connected stochastic Green's functions in the standard SQM
\bea
\label{eq:generating-functional-standard-eq1}
 Z( J_{\hat V}, J_{\hat \varpi} ) 
 & \equiv &
 e^{W ( J_{\hat V}, J_{\hat \varpi} ) }  
 \equiv
\int \!\!\! {\cal D}{\hat V}{\cal D}{\hat \varpi} e^{K ( {\hat V}, {\hat \varpi} ) + K_{\rm ex}}     \ , \nn 
K_{ex} 
 & \equiv & 
\int \!\!\! d^8z d\t {\rm Tr}
( J_{\hat V} {\hat V} + J_{\hat \varpi} {\hat \varpi} )      \ . 
\ena
Here the stochastic action $K ( {\hat V}, {\hat \varpi} )$ is defined in (\ref{eq:invariant-stochastic-action-eq2}). The one parameter family of covariant gauges is defined in terms of the Zwanziger's gauge fixing functions (\ref{eq:stochastic-gauge-fixing-eq1}).    
The effective stochastic action, which is a generator of the 1-P-I vertices in the connected stochastic Feynman diagrams, is defined by 
\bea
\label{eq:generating-functional-standard-eq2}
\Gamma ( {\overline V}, {\bar  \varpi} ) 
 =  W - \int\!\!\!\! d^8z d\t 
  {\rm Tr} ( J_{\hat V} {\overline V} + J_{\hat \varpi} {\bar \varpi} )       \ , \ 
{\overline V} 
 \equiv  \displaystyle{\frac{\delta W}{\delta J_{\hat V}}}   \ , \
{\bar \varpi} 
\equiv \displaystyle{\frac{\delta W}{\delta J_{\hat \varpi}}}   \ .
\ena
The stochastic Ward-Takahashi identity is expressed as $\Gamma ( {\bar V}, 0 ) = 0$.\cite{ZZ} It is also possible to derive the Ward-Takahashi identity for the BRST symmetry by introducing the BRST invariant stochastic action and the additional external sources coupled with the variations of the BRST transformation\cite{Nakazawa2} by applying the argument given for the YM case\cite{ZZ}. 

 We also define a generating functional of connected stochastic Green's functions in the stochastic BFM. 
\bea
\label{eq:generating-functional-background-eq1}
{\tilde Z}( J_V, J_\varpi, \Omega, \Omega^\dagger, {\bf \Pi} ) 
 & \equiv &
 e^{{\tilde W}( J_V, J_\varpi, \Omega, \Omega^\dagger, {\bf \Pi} ) }  
 \equiv
\int \!\!\! {\cal D}V{\cal D}\varpi 
e^{{\tilde K} ( V, \varpi, \Omega, \Omega^\dagger, {\bf \Pi} ) + {\tilde K}_{\rm ex}}     \ , \nn 
{\tilde K}_{ex} 
 & \equiv & 
\int \!\!\! d^8z d\t {\rm Tr}
( J_V V + J_\varpi \varpi )      \ . 
\ena
Here the stochastic action ${\tilde K} ( V, \varpi, \Omega, \Omega^\dagger, {\bf \Pi} )$ is defined by (\ref{eq:invariant-stochastic-action-eq3}). The background local gauge invariant stochastic gauge fixing is given by (\ref{eq:background-stochastic-gauge-fixing-eq1}). 
The corresponding effective stochastic action is defined by 
\bea
\label{eq:generating-functional-background-eq2}
{\tilde \Gamma} ( {\tilde V}, {\tilde \varpi}, \Omega, \Omega^\dagger, {\bf \Pi} ) 
 =  {\tilde W} - \int\!\!\!\! d^8z d\t {\rm Tr}( J_V {\tilde  V} + J_\varpi {\tilde \varpi} )         \ , \ 
{\tilde  V} 
 \equiv  \displaystyle{\frac{\delta {\tilde W}}{\delta J_V}}   \ , \
{\tilde  \varpi} 
\equiv \displaystyle{\frac{\delta {\tilde W}}{\delta J_\varpi}}   \ .
\ena
With these definitions, what we will show in the following is that the effective stochastic action in the stochastic BFM (\ref{eq:generating-functional-background-eq2}) is equivalent to the standard definition (\ref{eq:generating-functional-standard-eq2}),  
\bea
\label{eq:generating-functional-equivalence-eq1}
{\tilde \Gamma} ( 0, 0, \Omega = \Omega^\dagger = {\bf V}, {\bf \Pi} ) 
= 
\Gamma ( {\overline V}, {\overline \varpi}) 
\Big|_{
{\overline V}={\bf V}, 
{\overline \varpi}= \displaystyle{\frac{e^{g L_{\bf V}}- e^{-g L_{\bf V}}}{2g L_{\bf V}}}{\bf \Pi}  
}    \ ,
\ena
up to the difference of the Zwanziger's gauge fixing functions and the redefinition of the background ( or external ) auxiliary superfield. Namely, in the r.h.s. of (\ref{eq:generating-functional-equivalence-eq1}), the effective stochastic action is evaluated with the gauge fixing functions different from that given in (\ref{eq:stochastic-gauge-fixing-eq1}).

The stochastic action (\ref{eq:invariant-stochastic-action-eq3}) in the stochastic BFM is obtained from the original stochastic action (\ref{eq:invariant-stochastic-action-eq1}) except the Zwanziger's gauge fixing functions 
\bea
\label{eq:generating-functional-equivalence-eq2}
{\tilde K} \Big( V, \varpi, \Omega, \Omega^\dagger, {\bf \Pi} \Big)  
& = & 
K \Big( 
e^{2g {\hat V}},\  
{\hat \Pi} \equiv \displaystyle{\frac{2g L_{\hat V}}{1- e^{-2g L_{\hat V}}}}{\hat \varpi}  
 \Big)                      \ , \nn 
e^{2g {\hat V}} 
 \equiv  
e^{g\Omega}e^{2gV} e^{g\Omega^\dagger},  
& {} & 
\displaystyle{\frac{2g L_{\hat V}}{1- e^{-2g L_{\hat V}}}}{\hat \varpi}  
\equiv  
e^{g L_\Omega} {\bf \Pi} + 
e^{g L_\Omega} \displaystyle{\frac{2g L_V}{1- e^{-2g L_V}}} \varpi     \ .
\ena 
We write the non-linear transformations (\ref{eq:generating-functional-equivalence-eq2}) from the original superfields to those in the stochastic BFM as
\bea
\label{eq:generating-functional-equivalence-eq3}
{\hat V} 
& \equiv & 
f( V, {\bf V} ) 
=  
V + {\bf V} + ...  \  ,  \nn
{\hat \varpi} 
& \equiv & 
h( \varpi, V, {\bf \Pi}, {\bf V} ) 
= \varpi + {\bf \Pi} + ... \ . 
\ena
Here we have chosen the gauge 
$\Omega = \Omega^\dagger 
= {\bf V}$. 
We note that the functions, $f$ and $h$, satisfy
\bea
\label{eq:generating-functional-transformation-eq1}
f( 0, {\bf V} ) 
& = & 
{\bf V},\quad  f( V, 0 ) = V      \nn
h( 0, 0, {\bf \Pi}, {\bf V} ) 
& = & 
\displaystyle{\frac{e^{g L_{\bf V}}- e^{-g L_{\bf V}}}{2g L_{\bf V}}}{\bf \Pi},\quad  h( \varpi, V, 0, 0 ) = \varpi \ . 
\ena 
To show the equivalence, we introduce external source terms 
\bea
\label{eq:generating-functional-equivalence-eq4}
{\tilde K}'_{ex} 
& \equiv &
 \!\!\! 
\int \!\!\! d^8z d\t {\rm Tr}
\Big( 
J_V ( f( V, {\bf V} ) - f( 0, {\bf V} ))     \nn
& {} & \qquad\qquad\qquad\qquad 
+ J_\varpi ( h( \varpi, V, {\bf \Pi}, {\bf V} ) - h( 0, 0, {\bf \Pi}, {\bf V} ) ) 
\Big)     , 
\ena
and redefine the generating functional for the stochastic BFM
\bea
\label{eq:generating-functional-background-eq3}
{\tilde Z}'( J_V, J_\varpi, {\bf V}, {\bf \Pi} ) 
  \equiv 
 e^{ {\tilde W}'( J_V, J_\varpi, {\bf V}, {\bf \Pi} ) }  
 \equiv
\int \!\!\! {\cal D}V{\cal D}\varpi 
e^{{\tilde K} ( V, \varpi, \Omega= \Omega^\dagger= {\bf V}, {\bf \Pi} ) + {\tilde K}'_{\rm ex}}       \ .
\ena
By changing the integration variables with the inverse transformations of $f$ and $h$, 
$
V = f^{-1} ( {\hat V}, {\bf V} )
$
and 
$
\varpi = h^{-1} ( {\hat \varpi}, V = f^{-1} ( {\hat V}, {\bf V} ), {\bf \Pi}, {\bf V} )  
$
respectively, 
the stochastic action ${\tilde K} ( V, \varpi, \Omega= \Omega^\dagger= {\bf V}, {\bf \Pi} )$ becomes the original one $K( {\hat V}, {\hat \varpi} )$ except the Zwanziger's gauge fixing functions. The Jacobian due to the changes of the integration variables is trivial\cite{GGRS}.  The change of the integration variable may induce the following Zwanziger's gauge fixing functions in $K( {\hat V}, {\hat \varpi} )$ instead of the standard one (\ref{eq:stochastic-gauge-fixing-eq1}) 
\bea
\label{eq:stochastic-gauge-fixing-unusual-eq1}
{\hat \Phi}
 & \equiv &
 i \displaystyle{\frac{\xi}{4}}
e^{- g L_{\Omega^\dagger}}{\overline {\cal D}}^2 {\cal D}^2 
 f^{-1} ( {\hat V}, {\bf V} )
 = 
i \displaystyle{\frac{\xi}{4}}
e^{- g L_{\Omega^\dagger}}{\overline {\cal D}}^2 {\cal D}^2 
( {\hat V} - {\bf V} + ... ) \ , \nn
{\hat {\overline \Phi}}
 & \equiv &
- i \displaystyle{\frac{\xi}{4}}
e^{g L_\Omega}{\cal D}^2 {\overline {\cal D}}^2 
f^{-1} ( {\hat V}, {\bf V} )
 = 
- i \displaystyle{\frac{\xi}{4}}
e^{g L_\Omega}{\cal D}^2 {\overline {\cal D}}^2 
( {\hat V} - {\bf V} + ... )  . \
\ena
Since the time evolution of the local gauge invariant observables does not depend on the choice of the Zwanziger's gauge fixing functions in the standard description in (\ref{eq:generating-functional-standard-eq1}) and (\ref{eq:generating-functional-standard-eq2}), it may be safe to say that the expectation values of the local gauge invariant observables evaluated with these unusual Zwanziger's gauge fixing functions (\ref{eq:stochastic-gauge-fixing-unusual-eq1}) are equivalent to those in the standard one (\ref{eq:stochastic-gauge-fixing-eq1}). This means that the expectation values evaluated in (\ref{eq:generating-functional-standard-eq1}) with (\ref{eq:stochastic-gauge-fixing-unusual-eq1}) is independent from the background fields. 

Under the change of the integration variables, the external source term is simply reduced to 
\bea
\label{eq:generating-functional-background-eq4}
{\tilde K}'_{ex} 
 \equiv  
\int \!\!\! d^8z d\t {\rm Tr}
\Big( 
J_V ( {\hat V} - {\bf V} ) 
+ 
J_\varpi ( {\hat \varpi} - \displaystyle{\frac{e^{g L_{\bf V}}- e^{-g L_{\bf V}}}{2g L_{\bf V}}}{\bf \Pi} ) 
\Big)      \ .
\ena
In the limit of the vanishing external sources $J_V = J_\varpi = 0$, we find ${\tilde Z}' \rightarrow {\tilde Z}$. This means that the expectation values of local gauge invariant observables evaluated in ${\tilde Z}( 0, 0, {\bf V}, {\bf \Pi})$ in the stochastic BFM must be the same as those given by $Z( 0, 0 )$ with (\ref{eq:stochastic-gauge-fixing-unusual-eq1}), where $Z( 0, 0 )$ depends on the background fields only through the Zwanziger's gauge fixing functions (\ref{eq:stochastic-gauge-fixing-unusual-eq1}) which does not affect to the expectation values of the local gauge invariant observables. 

For ${\tilde Z}'$, we also introduce the effective stochastic action
\bea
\label{eq:generating-functional-background-eq5}
{\tilde \Gamma}' ( {\tilde V}', {\tilde \varpi}', {\bf V}, {\bf \Pi} ) 
 =  {\tilde W}' - \int\!\!\!\! d^8z d\t {\rm Tr}( J_V {\tilde  V}' + J_\varpi {\tilde \varpi}' )         \ , \ 
{\tilde  V}' 
 \equiv  \displaystyle{\frac{\delta {\tilde W}'}{\delta J_V}}   \ , \
{\tilde  \varpi}' 
\equiv \displaystyle{\frac{\delta {\tilde W}'}{\delta J_\varpi}}   \ .
\ena
By definition, we obtain
$
{\tilde  V}' 
 =  
f( {\tilde V}, {\bf V} ) - {\bf V}    \ , \
{\tilde \varpi}' 
 =  
h( {\tilde \varpi}, {\tilde V}, {\bf \Pi}, {\bf V} ) - \displaystyle{\frac{e^{g L_{\bf V}}- e^{-g L_{\bf V}}}{2g L_{\bf V}}}{\bf \Pi}    \ . 
$
Therefore ${\tilde V} = {\tilde \varpi} = 0$ implies ${\tilde V}' = {\tilde \varpi}' = 0$. While, keeping in mind the equivalence of ${\tilde Z}'$ to $Z$ with (\ref{eq:stochastic-gauge-fixing-unusual-eq1}),  we find 
\bea
\label{eq:generating-functional-transformation-eq2}
{\overline V}        
& = & 
f( {\tilde V}, {\bf V} )    \ ,     \nn 
{\bar \varpi} 
& = & 
h( {\tilde \varpi}, {\tilde V}, {\bf \Pi}, {\bf V} )        \ . 
\ena
Here we also note that ${\overline V}$ and ${\bar \varpi}$ are defined in (\ref{eq:generating-functional-standard-eq2}) with  (\ref{eq:stochastic-gauge-fixing-unusual-eq1}). 
Therefore, in the limit ${\tilde V} = {\tilde \varpi} = 0$, we have 
${\tilde \Gamma}' ( 0, 0, {\bf V}, {\bf \Pi} ) 
= {\tilde \Gamma} ( 0, 0, {\bf V}, {\bf \Pi} ) $, and 
\bea
\label{eq:generating-functional-equivalence-eq5}
{\tilde \Gamma}' ( 0, 0, {\bf V}, {\bf \Pi} ) 
& = & 
 W - \int\!\!\!\! d^8z d\t {\rm Tr}\Big( 
J_V {\bf  V} + J_\varpi \displaystyle{\frac{e^{g L_{\bf V}}- e^{-g L_{\bf V}}}{2g L_{\bf V}}}{\bf \Pi}    
\Big)    \ , \nn 
& = & 
\Gamma ( {\overline V}, {\bar \varpi} ) \Big|_{{\overline V}={\bf V},\ {\bar \varpi}= \displaystyle{\frac{e^{g L_{\bf V}}- e^{-g L_{\bf V}}}{2g L_{\bf V}}}{\bf \Pi} }       \ . 
\ena
 This shows the equivalence of the effective stochastic action in the stochastic BFM to the standard one with the gauge (\ref{eq:stochastic-gauge-fixing-unusual-eq1}) up to the redefinition of the external auxiliary superfield, 
 $
  {\bar \varpi}= \displaystyle{\frac{e^{g L_{\bf V}}- e^{-g L_{\bf V}}}{2g L_{\bf V}}}{\bf \Pi}
 $. 
 We note that, in this redefinition of the background auxiliary superfield, the difference of a factor $e^{g L_{\bf V}}$ in comparison with (\ref{eq:generating-functional-equivalence-eq2}) comes from the choice of the gauge $\Omega = \Omega^\dagger = {\bf V}$ as shown in (\ref{eq:background-vector-superfield-eq1}) and (\ref{eq:background-covariant-time-derivarive-A-eq3}). 
 
Here we have shown that the stochastic BFM in (\ref{eq:generating-functional-background-eq1}) is equivalent to the standard SQM (\ref{eq:generating-functional-standard-eq1}) with the unusual gauge fixing (\ref{eq:stochastic-gauge-fixing-unusual-eq1}).
In (\ref{eq:generating-functional-standard-eq1}), the probability distribution reproduces the standard Faddeev-Popov distribution in the equilibrium limit.  On the other hand, the conventional BFM for SYM$_4$ is equivalent to standard path-integral method. Therefore the proof explains the reason why the background covariant choice of the Zwanziger's gauge fixing functions (\ref{eq:background-stochastic-gauge-fixing-eq1}) simulates the contributions of the Nielsen-Kallosh ghost as well as the Faddeev-Popov ghost introduced in the conventional BFM. 

In the proof of the equivalence in this section, we have used the fact that the time evolution of the local gauge invariant observables does not depend on the stochastic gauge fixing procedure which implies that the expectation values of such observables evaluated in (\ref{eq:generating-functional-standard-eq1}) with (\ref{eq:stochastic-gauge-fixing-unusual-eq1}) are independent from the background field, because the background field dependence appears only in the stochastic gauge fixing functions (\ref{eq:stochastic-gauge-fixing-unusual-eq1}). One might suspect that the background field dependence of the stochastic gauge fixing procedure causes non-trivial background field dependence on the expectation values. In order to confirm the background field independence, in the next section, we discuss this issue in view of the BRST invariant formulation of the SQM approach.

%\newpage
%%%%%%%%%%%%%%%%%%%%%%%%%%%%%%%%%%%%%%%%%%%%%%%%%%%%%%%%%%%%%%%%%%%%%%%%%%%%%%
%2003.12.3
%Added in the revised version for the publication in Nuclear Physics B
%%%%%%%%%%%%%%%%%%%%%%%%%%%%%%%%%%%%%%%%%%%%%%%%%%%%%%%%%%%%%%%%%%%%%%%%%%%%%%

\section{BRST symmetry in the SQM approach and the background field independence of the expectation values of observables}

In the last of this note, we add a brief discussion on the BRST symmetry in the stochastic BFM. At first, we recall the following fact. In the standard path-integral approach, the existence of the BRST symmetry is necessary in a gauge theory for its unitarity of the S-matrix as well as its renormalizability, though it is not a sufficient condition. This is also true for the conventional BFM. The S-matrix elements can be constructed from the effective action (i.e., the generator of the 1-P-I vertices) in the conventional BFM.\cite{Boulware}\cite{Hart}\cite{AGS} The BRST symmetry is also useful to prove the renormalizability\cite{Grassi} and the background gauge equivalence\cite{BC} of the effective action rigorously in this context. By contrast, in the standard SQM applied to the gauge theory, the stochastic action $K$ and the effective stochastic action $\Gamma$ defined in (\ref{eq:generating-functional-standard-eq2}) are not directly related to the S-matrix. The BRST symmetry in terms of the stochastic action ensures its renormalizability, however, the unitarity is not directly related with the BRST symmetry of the stochastic action. It is the stochastic gauge fixing procedure that assures the unitarity.\cite{Zwanziger}\cite{BZ}\cite{NOOY}\cite{BHST} In a perturbative sense, this comes from the decoupling of the FP ghost, which is introduced in the BRST invariant formulation of the standard SQM approach, generally occured in the so-called \lq\lq flow gauges\rq\rq that is a non-covariant gauge in the sense of the 5-dimensional \lq\lq Lorentz covariance\rq\rq.\cite{CH}\cite{ZZ}\cite{Nakazawa3} The FP ghost does not contribute to the renormalization of the gauge field sector. In spite of this decoupling of the FP ghost, the BRST symmetry is useful in applying the field theoretical techniques, such as the arguments in terms of the Ward-Takahashi identities, to the SQM approach.\cite{ZZ} In this respect, although it is convenient to respect the background local gauge invariance in the stochastic BFM for the perturbative analysis of SYM$_4$, it is worthwhile to discuss the BRST symmetry in this context. 

For YM case, the BRST invariant formulation of SQM\cite{ZZ}\cite{KOT}\cite{Nakazawa3} has been used in Ref.\cite{MOSZ} to prove the equivalence of the stochastic BFM to the standard one. We extend this argument to SYM$_4$ and verify the background field independence of the correlation functions or expectation values of local gauge invariant observables evaluated in the standard SQM approach with the unusual gauge fixing (\ref{eq:stochastic-gauge-fixing-unusual-eq1}) in the BRST invariant formulation. It is shown that the partition function defined in the stochastic BFM for SYM$_4$ is equivalent to the standard one up to a BRST exact term in view of their BRST invariant formulations. This also completes the proof given in the previous section. 

In the SQM approach for SYM$_4$, the BRST symmetry is defined from the 5-dimensional local gauge symmetry (\ref{eq:extended-local-gauge-transf-eq3}).\cite{Nakazawa2} The consistent truncation of this 5-dimensional BRST symmetry is carried out by integrating the auxiliary superfields, $\Phi$ and ${\bar \Phi}$, and Nakanishi-Lautrup superfields. Then, we obtain a truncated BRST symmetry in an extended phase space 
$\Big\{ ( {\hat V}^a, {\hat \varpi}_a ), ( {\hat c}^a, {\hat {c'}}^a ), ( {\hat {\bar c}}^a, {\hat {\bar c}}^{'a} ) \Big\} $. It is defined by 
\bea
\label{eq:stochastic-BRST-transf-eq1}
{\hat \delta}_{\rm BRST} {\hat V}^a 
& = & 
- \lambda \displaystyle{\frac{i}{2}} {\hat L}^a_{\ b}{\hat c}^b 
+ \lambda \displaystyle{\frac{i}{2}} {\hat {\bar c}}^b {\hat L}_b^{\ a}   \ , \nn 
{\hat \delta}_{\rm BRST} {\hat \varpi}_a 
& = & 
 \lambda \displaystyle{\frac{i}{2}} \pa_a {\hat L}^c_{\ b} {\hat \varpi}_c {\hat c}^b 
- \lambda \displaystyle{\frac{i}{2}} {\hat {\bar c}}^b \pa_a {\hat L}_b^{\ c} {\hat \varpi}_c  
 \ , \nn 
{\hat \delta}_{\rm BRST} {\hat c}^a
& = & 
- \lambda \displaystyle{\frac{g}{2}} [{\hat c}\times {\hat c}]^a           \ , \nn
{\hat \delta}_{\rm BRST} {\hat {\bar c}}^a   
& = & 
- \lambda  \displaystyle{\frac{g}{2}} [{\hat {\bar c}}\times {\hat {\bar c}}]^a \ , \nn
{\hat \delta}_{\rm BRST} {{\hat c}'}_a   
& = & 
\lambda (-\displaystyle{\frac{1}{4}} {\bar D}^2 ) ( \displaystyle{\frac{i}{2}} {\hat \varpi}_b {\hat L}^b_{\ a}  )
- \lambda g [{\hat c}\times {\hat c}']_a           \ , \nn 
{\hat \delta}_{\rm BRST} {{\hat {\bar c}}'}_a
& = &  
- \lambda (-\displaystyle{\frac{1}{4}} D^2 ) ( \displaystyle{\frac{i}{2}}  {\hat L}_a^{\ b} {\hat \varpi}_b ) 
 - \lambda g [{\hat {\bar c}}\times {\hat {\bar c}}']_a      
     \ .       
\ena
We note that this truncated BRST transformation is $nilpotent$. It preserves the chirality of the superfields: 
${\bar D} {\hat c} = {\bar D} {\hat c}' = D {\hat {\bar c}} = D {\hat {\bar c}}' = 0$. The truncated BRST invariant stochastic action is reduced from the 5-dimensional extended one as 
\bea
\label{eq:BRST-invariant-stochastic-action-eq1}
Z_{\rm BRST} 
& = &  
\int{\cal D}{\hat V}{\cal D}{\hat \varpi}{\cal D}{\hat c}{\cal D}{\hat c}'{\cal D}{\hat {\bar c}}{\cal D}{{\hat {\bar c}}'} e^{ K_{\rm BRST}}        \ , \nn
K_{\rm BRST} 
& \equiv & 
K_0 + i \int \!\!\! d^6z d\t  
 {{\hat c}'}_a {\dot {\hat c}}^a 
+ i \int \!\!\! d^6{\bar z} d\t {{\hat {\bar c}}'}_a {\dot {\hat {\bar c}}}^a   \nn 
& {} & \qquad\qquad 
+  \displaystyle{\frac{\xi}{2\kappa}}\int \!\!\! d^8z d\t  {\hat \delta}'_{\rm BRST} ( - {{\hat {\bar c}}'}_a {\bar D}^2 {\hat V}^a  +  {{\hat c}'}_a D^2 {\hat V}^a )       \ .
\ena
Here ${\hat \delta}_{\rm BRST} \equiv \lambda {\hat \delta}'_{\rm BRST}$, $d^6z = d^4x d^2{\theta}$ and  $d^6{\bar z} = d^4x d^2{\bar \theta}$. $K_0$ is defined by the stochastic action (\ref{eq:invariant-stochastic-action-eq2}) except the stochastic gauge fixing term, which is included in the BRST exact part in (\ref{eq:BRST-invariant-stochastic-action-eq1}). The BRST invariance of this truncated stochastic action $K_{\rm BRST}$ is manifest because of the invariance of the derivative terms with repect to the stochastic time:  
$
\delta_{\rm BRST} \Big\{ 
\int \!\! d^8z d\t {\hat \varpi}_a {\dot {\hat V}}^a  
+  
\int \!\! d^6z d\t {{\hat c}'}_a {\dot {\hat c}}^a 
+ 
 \int \!\! d^6{\bar z} d\t {{\hat {\bar c}}'}_a {\dot {\hat {\bar c}}}^a 
 \Big\} 
= 0 . 
$ 
Since (\ref{eq:BRST-invariant-stochastic-action-eq1}) can be regarded as a Legendre transformation, the canonical conjugate momentum of the FP ghost ${\hat c}$ (${\hat {\bar c}}$) with respect to the stochastic time is ${\hat c}'$ (${\hat {\bar c}}'$). 

The expression (\ref{eq:BRST-invariant-stochastic-action-eq1}) is useful to discuss the possibility of another choice of the Zwanziger's gauge fixing functions. For example, we may choose the gauge fixing functions, 
$
{\hat \Phi} 
 = i {\xi\over 4}
 {\overline D}^2 {\hat {\cal D}}^2 {\hat V} 
$
 and 
$ 
{\hat {\bar \Phi}}
 = - i {\xi\over 4} 
 D^2 {\hat {\overline {\cal D}}}^2 {\hat V}   
$
, which are allowed on dimensional ground. Furthermore, the weak coupling limit  $g \rightarrow 0$ is equivalent to (\ref{eq:stochastic-gauge-fixing-eq1}). For these gauge fixing functions, the BRST exact term in (\ref{eq:BRST-invariant-stochastic-action-eq1}) becomes 
$
 {\xi\over 2\kappa}\int \!\! d^8z d\t  {\hat \delta}'_{\rm BRST} ( - {{\hat {\bar c}}'}_a {\hat {\bar {\cal D}}}^2 {\hat V}^a  +  {{\hat c}'}_a {\hat {\cal D}}^2 {\hat V}^a )   
$ . 
Therefore, though the gauge fixing term generates the new interactions, we conclude that the correlation functions of the local gauge invariant (i.e.,  BRST invariant) observables are the same as those evaluated in (\ref{eq:stochastic-gauge-fixing-eq1}). 

The superpropagators for vector superfields ${\hat V}$ and ${\hat \varpi}$ are given in (\ref{eq:super-propagator-eq1}). For the Faddeev-Popov ghosts, we obtain
\bea
\label{eq:super-propagator-eq2}
\langle {\hat c}^a ( \t, z ) {\hat c}^{' b} ( \t', z' ) \rangle      
& = & 
- i 
\displaystyle{\frac{1}{(  i\omega + k^2/\kappa )}} 
(- \displaystyle{\frac{1}{4}} {\tilde {\bar D}}^2 )
\delta^2 ( \theta - \theta' ) \delta^2 ( {\bar \theta} - {\bar \theta'} )            \ , \nn
\langle {\hat {\bar c}}^a ( \t, z ) {\hat {\bar c}}^{'b} ( \t', z' ) \rangle
& = & 
- i 
\displaystyle{\frac{1}{(  i\omega + k^2/\kappa )}} 
(- \displaystyle{\frac{1}{4}} {\tilde D}^2 )
\delta^2 ( \theta - \theta' ) \delta^2 ( {\bar \theta} - {\bar \theta'} )            \ .
\ena
Here, we have suppressed the momentum integration 
$\int \!\! d^4k d\omega (2\pi)^{-5}e^{ik \cdot ( x - x' ) + i \omega ( \t - \t' )}$. 
We have also defined the covariant derivatives 
${\tilde D}_\a \equiv \pa_\a - \sigma^m_{\a{\dot \a}} \theta^{\dot \a}k_m $ 
and 
${\tilde {\bar D}}_{\dot \a} \equiv - {\bar \pa}_{\dot \a} + \theta_\a \sigma^m_{\a{\dot \a}} k_m $. 
We note that the superpropagators for FP ghosts are retarded one. In the perturbative analysis, as we have already mentioned on the FP ghost decoupling in the stochastic gauge fixing procedure, this means that the FP ghost loops do not contribute to the renormalization of the vector superfields sector, $K_0$ in (\ref{eq:BRST-invariant-stochastic-action-eq1}). 

The BRST invariant formulation in terms of the stochastic action principle is renormalizable which can be shown by the arguments based on the power counting and the Ward-Takahashi identities, which are derived from the Parisi-Sourlas type hidden supersymmetry in (\ref{eq:Parisi-Sourlas-eq2}), the BRST symmetry (\ref{eq:stochastic-BRST-transf-eq1}) and the scale invariance of the FP ghosts
\bea
\label{eq:FP-ghost-number-eq1}
{\hat c} 
& \rightarrow & e^{\rho} {\hat c}    \ , \quad 
{\hat c}' 
 \rightarrow  e^{-\rho} {\hat c}'     , \nn
{\hat {\bar c}} 
& \rightarrow & e^{\rho} {\hat {\bar c}}    \ , \quad 
{\hat {\bar c}}' 
 \rightarrow  e^{- \rho} {\hat {\bar c}}'     . 
\ena
The power counting of the stochastic Feynman diagrams generated by (\ref{eq:BRST-invariant-stochastic-action-eq1}) is similar to what we have explained in (\ref{eq:power-counting-eq3}). 
The degree of the divergence is given by 
\bea
\label{eq:power-counting-eq4} 
4 - 2E_\varpi - \displaystyle{\frac{3}{2}} 
( E_c + E_{c'} + E_{\bar c} + E_{{\bar c}'} )   \ , 
\ena
where $E_\varpi$, $E_c$, $E_{c'}$, $E_{\bar c}$ and $E_{{\bar c}'}$ are the numbers of the external auxiliary vector superfield and the FP ghosts: $\varpi$, $c$, $c'$, ${\bar c}$ and ${\bar c}'$, respectively. 
Therefore, the stochastic action (\ref{eq:BRST-invariant-stochastic-action-eq1}) is multiplicatively renormalizable by the renormalizations of the wave functions and the coupling constants, $g$
 and $\kappa$.\cite{Nakazawa2}  

It is straightforward to define the BRST symmetry in the stochastic BFM. We make changes of the integration variables in (\ref{eq:BRST-invariant-stochastic-action-eq1}) by (\ref{eq:generating-functional-equivalence-eq3}) (or equivalently, (\ref{eq:generating-functional-equivalence-eq2})) for ${\hat V}$ and ${\hat \varpi}$. The changes of the integration variables for the FP ghosts in (\ref{eq:BRST-invariant-stochastic-action-eq1}) are also carried out by 
\bea
\label{eq:change-integration-variable-eq1}
{\hat c} 
& \rightarrow&  e^{-g\Omega^\dagger} c e^{g\Omega^\dagger}                  \ , \qquad 
{\hat {\bar c}} 
\rightarrow e^{g\Omega}{\bar c} e^{-g\Omega}    \ ,  \nn
{\hat c'} 
& \rightarrow&  e^{-g\Omega^\dagger} c' e^{g\Omega^\dagger}                  \ , \qquad 
{\hat {\bar c}}' 
\rightarrow e^{g\Omega}{\bar c}' e^{-g\Omega}    \ . 
\ena 
Namely, the chiral FP ghost superfields ${\hat c}$ and ${\hat c}'$ are replaced to the background chiral superfields $c$ and $c'$, and the anti-chiral FP ghost superfields ${\hat {\bar c}}$ and ${\hat {\bar c}}'$ are replaced to the background anti-chiral superfields ${\bar c}$ and ${\bar c}'$. This is a consequence of the quantum local gauge transformation (\ref{eq:quantum-local-gauge-transf-eq2}). 
We also note that we do not consider the background fields for the FP ghosts, ${\hat c}$, ${\hat c}'$, ${\hat {\bar c}}$ and ${\hat {\bar c}}'$. 
For the definition of the stochastic BFM, in addition to these changes of the integration variables, the stochastic gauge fixing term must be specified so as to be invariant under the background local gauge transformation. 
 
 The BRST transformation for the stochastic BFM is defined by 
\bea
\label{eq:stochastic-BRST-transf-eq2}
\delta_{\rm BRST} V^a 
& = & 
- \lambda \displaystyle{\frac{i}{2}} L^a_{\ b}c^b 
+ \lambda \displaystyle{\frac{i}{2}} {\bar c}^b L_b^{\ a}   \ , \nn 
\delta_{\rm BRST} \varpi_a 
& = & 
 \lambda \displaystyle{\frac{i}{2}} \pa_a L^c_{\ b} \varpi_c c^b 
- \lambda \displaystyle{\frac{i}{2}} {\bar c}^b \pa_a L_b^{\ c} \varpi_c  
 \ , \nn 
\delta_{\rm BRST} c^a
& = & 
- \lambda \displaystyle{\frac{g}{2}} [c\times c]^a           \ , \nn
\delta_{\rm BRST} {\bar c}^a   
& = & 
- \lambda  \displaystyle{\frac{g}{2}} [{\bar c}\times {\bar c}]^a \ , \nn
\delta_{\rm BRST} {c'}_a   
& = & 
\lambda (-\displaystyle{\frac{1}{4}} {\bar {\cal D}}^2 )  \displaystyle{\frac{i}{2}} \Big( 
( e^{-2g L_V} {\bf \Pi} )_a + \varpi_b L^b_{\ a}  
\Big) 
- \lambda g [c\times c']_a           \ , \nn 
\delta_{\rm BRST} {{\bar c}'}_a
& = &  
- \lambda (-\displaystyle{\frac{1}{4}} {\cal D}^2 )  \displaystyle{\frac{i}{2}}  ( {\bf \Pi}_a + L_a^{\ b} \varpi _b ) 
 - \lambda g [{\bar c}\times {\bar c}']_a      
     \ .       
\ena
This is deduced from the 5-dimensional BRST transformation by the consistent truncation with the background-quantum splitting in the stochastic BFM. 
We also define the BRST transformation of the background fields. The background fields $\Omega$ and $\Omega^\dagger$ are invariant while ${\bf \Pi}$ is transformed as an adjoint matter by (\ref{eq:quantum-local-gauge-transf-eq1}) under the quantum type local gauge transformation. Thus we define 
\bea
\label{eq:stochastic-BRST-transf-eq3}
\delta_{\rm BRST} \Omega_a 
& = & 
\delta_{\rm BRST} \Omega^\dagger_a = 0  \ , \nn
\delta_{\rm BRST} {\bf \Pi}_a 
& = & - \lambda g [ {\bar c} \times {\bf \Pi}]_a    \ . 
\ena
The BRST transformation, (\ref{eq:stochastic-BRST-transf-eq2}) and (\ref{eq:stochastic-BRST-transf-eq3}), is also $nilpotent$. It preserves the chirality of the superfields with respect to the background field: 
${\bar {\cal D}} c = {\bar {\cal D}} c' = {\cal D} {\bar c} = {\cal D} {\bar c}' = 0$. The corresponding BRST invariant stochastic action is given by 
\bea
\label{eq:BRST-invariant-stochastic-action-eq2}
{\tilde Z}_{\rm BRST} 
& = &  
\int{\cal D}V{\cal D}\varpi{\cal D}c{\cal D}c'{\cal D}{\bar c}{\cal D}{{\bar c}'} e^{ {\tilde K}_{\rm BRST}}        \ , \nn
{\tilde K}_{\rm BRST} 
& \equiv & 
{\tilde K}_0 + i \int \!\!\! d^6z d\t  
 {c'}_a {\bar {\cal D}}_\t c^a 
+ i \int \!\!\! d^6{\bar z} d\t {{\bar c}'}_a {\cal D}_\t {\bar c}^a   \nn 
& {} & \qquad\qquad 
+  \displaystyle{\frac{\xi}{2\kappa}}\int \!\!\! d^8z d\t  \delta'_{\rm BRST} ( - {{\bar c}'}_a {\bar {\cal D}}^2 V^a  +  {c'}_a {\cal D}^2 V^a )       \ .
\ena
Here $\delta_{\rm BRST} \equiv \lambda \delta'_{\rm BRST}$. ${\tilde K}_0$ is defined by (\ref{eq:invariant-stochastic-action-eq3}) except the stochastic gauge fixing term, which is included in the BRST exact term in (\ref{eq:BRST-invariant-stochastic-action-eq2}). For the BRST invariance of ${\tilde K}_{\rm BRST}$ in (\ref{eq:BRST-invariant-stochastic-action-eq2}), we note 
\bea
\label{eq:BRST-invariance-eq1}
& {} & 
\delta_{\rm BRST} \Big\{ 
\int \!\! d^8z d\t ( K_a^{\ b}{\bf \Pi}_b + \varpi_a  
)\Big( 
{\dot V}^a  + \displaystyle{\frac{1}{2g}} L_b^{\ a}{\cal D}^{' b}_\t - \displaystyle{\frac{1}{2g}} L^a_{\ b} {\bar {\cal D}}^{' b}_\t
\Big)       \nn 
& {} & \qquad\qquad  
+  
\int \!\! d^6z d\t {c'}_a {\bar {\cal D}}_\t  c^a 
+ 
 \int \!\! d^6{\bar z} d\t {{\bar c}'}_a {\cal D}_\t {\bar c}^a 
 \Big\} 
= 0    \ . 
\ena

The BRST exact part in (\ref{eq:BRST-invariant-stochastic-action-eq2}) is decomposed as 
\bea
\label{eq:FP-ghost-sector-eq1}
& {} & 
\displaystyle{\frac{\xi}{2\kappa}} \int \!\!\! d^8z d\t  \delta'_{\rm BRST} ( - {{\bar c}'}_a {\bar {\cal D}}^2 V^a  +  {c'}_a {\cal D}^2 V^a )   \nn
& {} & \quad
= 
- i \displaystyle{\frac{\xi}{16\kappa}} \int \!\!\! d^8z d\t  
 \varpi_a  \Big( 
L_b^{\ a} {\cal D}^2{\bar {\cal D}}^2 V^b + L^a_{\ b} {\bar {\cal D}}^2 {\cal D}^2 V^b 
   \Big)     \nn 
& {} & \qquad 
 - i \displaystyle{\frac{\xi}{16\kappa}} \int \!\!\! d^8z d\t  
 {\bf \Pi}_a  \Big( 
 {\cal D}^2{\bar {\cal D}}^2 V^a + ( e^{2g L_V} {\bar {\cal D}}^2 {\cal D}^2 V     )^a 
   \Big)     \nn   
& {} & \qquad 
+ \displaystyle{\frac{\xi}{2\kappa}} \int \!\!\! d^8z d\t    
\Big\{ 
- \displaystyle{\frac{i}{2}} ({\bar c}'_a {\bar {\cal D}}^2 - c'_a {\cal D}^2 ) ( L_{\ b}^a c^b - {\bar c}^b L_b^{\ a} )    \nn
& {} & \qquad \qquad \qquad \qquad 
+ g [{\bar c}\times {\bar c}']_a {\bar {\cal D}}^2 V^a 
-g [c\times c']_a {\cal D}^2 V^a         \Big\}       \ . 
\ena
The first and the second terms in the r.h.s. of this expression reproduce the background local gauge invariant stochastic gauge fixing terms in (\ref{eq:invariant-stochastic-action-eq3}) with (\ref{eq:background-stochastic-gauge-fixing-eq1}). The third term describes the interactions between the vector superfield and the FP ghosts. The interaction terms in the BRST exact part do not contribute to the renormalization of the vector superfields sector, while these terms are renormalized by the perturbative correction due to the vector superfields. The power counting argument is essentially the same as (\ref{eq:power-counting-eq4}) which assures the renormalizability of the BRST invariant formulation of the stochastic BFM. 
We comment on the difference between the stochastic BFM and the conventional BFM. In the conventional BFM, we need the Nielsen-Kallosh ghost. By contrast, in the stochastic BFM in the BRST invariant formulation, it is not necessary to introduce an additional ghost corresponding to the Nielsen-Kallosh ghost. This results from that we do not introduce a device such as the quadratic term of the Nakanishi-Lautrup fields, which are chiral and anti-chiral, in defining the stochastic gauge fixing term in (\ref{eq:BRST-invariant-stochastic-action-eq1}) and also in (\ref{eq:BRST-invariant-stochastic-action-eq2}). 

We also comment that the Jacobian under the changes of the integration variables to define the stochastic BFM, (\ref{eq:generating-functional-equivalence-eq3}) and (\ref{eq:change-integration-variable-eq1}), is trivial. We have already mentioned this fact for the vector superfields $V$ and $\varpi$ in the previous section. For the FP-ghosts, it is a consequence of the retarded nature of the superpropagators of the FP ghosts. Their superpropagators in the Feynman gauge are given by
\bea
\label{eq:super-propagator-eq3}
\langle c^a ( \t, z ) c^{' b} ( \t', z' ) \rangle      
& = & 
- i 
\Big(
{\bar {\cal D}}_\t -  
\displaystyle{\frac{{\bar {\cal D}}^2{\cal D}^2}{16\kappa}} 
\Big)^{-1}
(- \displaystyle{\frac{1}{4}} {\bar {\cal D}}^2 )
\delta^8 ( z-z' ) \delta ( \t-\t' )           \ , \nn
\langle {\bar c}^a ( \t, z ) {\bar c}^{'b} ( \t', z' ) \rangle
& = & 
- i 
\Big(
{\cal D}_\t -  
\displaystyle{\frac{{\cal D}^2{\bar {\cal D}}^2}{16\kappa}} 
\Big)^{-1}
(- \displaystyle{\frac{1}{4}} {\cal D}^2 )
\delta^8 ( z-z' ) \delta ( \t-\t' )          \ , 
\ena
in the vanishing limit of the quantum vector superfield $V \rightarrow 0$. 
Because of the retarded nature of these superpropagators, the FP-ghost loops do not contribute to relevant stochastic Feynman diagrams. This means that the Jacobian of the changes of the integration variables for the FP ghosts, (\ref{eq:change-integration-variable-eq1}), is trivial, even if it could be evaluated by, for example, a gaussian cut off regulator with the differential operators in (\ref{eq:super-propagator-eq3}).

Now we discuss the equivalence of the stochastic BFM (\ref{eq:BRST-invariant-stochastic-action-eq2}) to the standard SQM (\ref{eq:BRST-invariant-stochastic-action-eq1}) in the context of the BRST invariant formulations. Let us consider the partition function which is defined by the changes of the integration variables in ${\tilde Z}_{\rm BRST}$ in (\ref{eq:BRST-invariant-stochastic-action-eq2}) by the inverse transformations of (\ref{eq:generating-functional-equivalence-eq3}) and (\ref{eq:change-integration-variable-eq1}). This procedure yields a relation between ${\tilde Z}_{\rm BRST}$ and $Z_{\rm BRST}$ in (\ref{eq:BRST-invariant-stochastic-action-eq1}). By definition $Z_{\rm BRST}$ is background field independent. 
 The possible background field dependence in ${\tilde Z}_{\rm BRST}$, after the changes of the integration variables may come from the background local gauge invariant stochastic gauge fixing procedure, since the Jacobian of these changes of the integration variables is trivial. The BRST transformation (\ref{eq:stochastic-BRST-transf-eq2}) is also converted to the orginal one (\ref{eq:stochastic-BRST-transf-eq1}). As we have already studied, we obtain the unusual stochastic gauge fixing term (\ref{eq:stochastic-gauge-fixing-unusual-eq1}).  Therefore, after the changes of the integration variables, ${\tilde Z}_{\rm BRST}$ becomes 
 ${\tilde Z}'_{\rm BRST}$,  
\bea
\label{eq:BRST-BFM-equivalence-eq1}
{\tilde Z}'_{\rm BRST}          
& \equiv & 
\int{\cal D}{\hat V}{\cal D}{\hat \varpi}{\cal D}{\hat c}{\cal D}{\hat c}'{\cal D}{\hat {\bar c}}{\cal D}{{\hat {\bar c}}'} e^{ {\tilde K}'_{\rm BRST}}        \ , \nn
{\tilde K}'_{\rm BRST} 
& \equiv & 
K_0 ({\hat V}, {\hat \varpi}) + i \int \!\!\! d^6z d\t  
 {{\hat c}'}_a {\dot {\hat c}}^a 
+ i \int \!\!\! d^6{\bar z} d\t {{\hat {\bar c}}'}_a {\dot {\hat {\bar c}}}^a  
+  {\hat \delta}'_{\rm BRST} X          \ , \nn
X 
& = & 
\displaystyle{\frac{\xi}{2\kappa}} {\rm Tr}\!\!\! \int \!\!\! d^8z d\t \Big\{ 
 {{\hat c}'}e^{-g L_{\Omega^\dagger}} \Big( {\cal D}^2 f^{-1}({\hat V}, {\bf V} )  \Big) 
 - {{\hat {\bar c}}'} e^{g L_\Omega}\Big( {\bar {\cal D}}^2 f^{-1}({\hat V}, {\bf V} )  \Big)   
  \Big\}   .
\ena
Here, ${\tilde K}'_{\rm BRST}$ includes the unusual stochastic gauge fixing (\ref{eq:stochastic-gauge-fixing-unusual-eq1}) in its BRST exact part and the difference of ${\tilde K}'_{\rm BRST}$ from the standard $K_{\rm BRST}$ is ${\rm BRST}$ $exact$. This implies the equivalence of the stochastic BFM to the standard SQM up to a BRST exact term in these BRST invariant formulations. Clearly, the background field dependence of ${\tilde Z}'_{\rm BRST}$ takes a BRST exact form. This indicates that ${\tilde Z}'_{\rm BRST}$ reproduces the background field $independent$ expectation values for the local gauge invariant (i.e., BRST invariant) observables. The generating functional of the correlation functions of a composite opetator ${\cal O}$ may be defined by 
\bea 
{\tilde Z}'_{\rm BRST} (J, I, {\cal J}, {\cal I}; \Omega, \Omega^\dagger) 
& = & 
\int \!\!\! {\cal D}Q e^{{\tilde K}'_{\rm BRST} + K_{\rm source}}    \ , \nn
K_{\rm source} 
& = & 
\int\!\!\! d^8z d\t 
 {\rm Tr}\Big\{ 
 \sum_M \Big( J_M Q^M + I_M ({\hat \delta}'_{\rm BRST} Q^M ) \Big)    \nn 
& {} & 
\qquad \qquad \qquad \qquad 
 + {\cal J} {\cal O} + {\cal I} ( {\hat \delta}'_{BRST} {\cal O}  )  
 \Big\}       \    .  
\ena
Here $Q^M$ denotes all the superfields, 
$Q^M 
\equiv ({\hat V}, {\hat \varpi}, {\hat c}, {\hat c}', {\hat {\bar c}}, {\hat {\bar c}}')$. $J_M$ denotes the corresponding external source currents. 
Since the background field dependence of ${\tilde K}'_{\rm BRST}$ is BRST exact, the variation with respect to the background fields is given by   
\bea
\delta_{\Omega, \Omega^\dagger} {\tilde Z}'_{\rm BRST} 
& = & 
\int\!\!\! {\cal D}Q \Big\{ 
{\hat \delta}'_{\rm BRST}( \delta_{\Omega, \Omega^\dagger} X ) 
\Big\} e^{{\tilde K}'_{\rm BRST} + K_{\rm source}}      \nn
& = & 
\int\!\!\! {\cal D}Q ( \delta_{\Omega, \Omega^\dagger} X ) \!\! 
\int\!\!\! d^8z d\t \sum_M ({\hat \delta}'_{\rm BRST} Q^M) \displaystyle{\frac{\delta e^{{\tilde K}'_{\rm BRST} + K_{\rm source}}}{\delta Q^M (\t, z)}}    \nn
& = & 
\int\!\!\! d^8z d\t \Big\{       \sum_M J_M \displaystyle{\frac{\delta}{\delta I_M}}  
+ {\cal J}\displaystyle{\frac{\delta}{\delta {\cal I}}} \Big\} 
\!\! \int\!\!\! {\cal D}Q ( \delta_{\Omega, \Omega^\dagger} X ) e^{{\tilde K}'_{\rm BRST} + K_{\rm source}}      , 
\ena
thereby vanishing in the limit $J_M \rightarrow 0$ for the local gauge invariant (i.e. BRST invariant) observable ${\hat \delta}'_{\rm BRST}{\cal O} = 0$. Therefore ${\tilde Z}'_{\rm BRST}$, in fact, generates the background field independent correlation functions for the local gauge invariant observables. This completes the proof given in the previous section.

%\newpage
%%%%%%%%%%%%%%%%%%%%%%%%%%%%%%%%%%%%%%%%%%%%%%%%%%%%%%%%%%%%%%%%%%%%%%%%%%%%%%
%
%
%%%%%%%%%%%%%%%%%%%%%%%%%%%%%%%%%%%%%%%%%%%%%%%%%%%%%%%%%%%%%%%%%%%%%%%%%%%%%%

\section{Discussions}

In this note, we have applied the stochastic BFM to SYM$_4$. By the explicit one-loop calculation, we obtain the one-loop $\beta$-function for the gauge coupling constant. The $\beta$-function agrees with that given in the conventional approach. Therefore we have confirmed by the explicit one-loop calculation that the stochastic gauge fixing procedure for SYM$_4$ is equivalent to the Faddeev-Popov prescription in the path-integral method. This is consistent to our previous formal proof on this equivalence\cite{Nakazawa2} in the following sense. 

%This confirm our proposal on SQM approach to SYM$_4$ in which the probability 
%distribution in the conventional approach is reproduced in a specified stochast%ic gauge fixing procedure. 

In the standard SQM approach to SYM$_4$, the stochastic gauge fixing procedure is equivalent to the Faddeev-Popov prescription in the superfield formalism. In this case, the necessary ghost is only the Faddeev-Popov ghost in the path-integral approach. The BRST invariant structure is also introduced in the standard stochastic action\cite{Nakazawa2} which ensures the perturbative renormalizability of SYM$_4$ in SQM. 

In contrast, the Faddeev-Popov prescription in the conventional BFM requires the Nielsen-Kallosh type ghost in addition to the Faddeev-Popov ghost in the superfield formalism. This is the particular feature of the conventional BFM in the superfield formalism. Our proposal for the stochastic gauge fixing procedure in the stochastic BFM is to choose the Zwanziger's gauge fixing functions to be invariant under the background local gauge transformation. The one-loop calculation in the stochastic BFM agrees with that in the conventional approach which supports that the backgound local gauge invariant stochastic gauge fixing procedure simulates the contributions of the Nielsen-Kallosh ghost as well as the Faddeev-Popov ghost. 

We have given a proof on the equivalence of the stochastic BFM to the standard one in the context of SQM by mapping the stochastic gauge fixing procedure defined in BFM into the standard SQM. 
In particular, the background field independence has been verified for the expectation values of the local gauge invariant observables, evaluated in an unusual background field dependent stochastic gauge fixing, in terms of the BRST invariant formulation of SQM. 
An important check on 
the equivalence may be the calculation of anomalies in SYM$_4$. For example, the superconformal anomaly in SYM$_4$ is known to be proportional to the $\beta$-function of the gauge coupling\cite{CPS}\cite{PiSi}\cite{GW-GMZ}. By deriving such an anomaly without using the relation to the $\beta$-function, we can also check that the stochastic BFM reproduces the contribution of both ghosts in the conventional BFM. 

In the context of the stochastic BFM, the superfield Langevin equation is given by  
\bea
\label{eq:Langevin-conclusion} 
& {} &  
 \displaystyle{\frac{d}{d\t}} V^a     
- 
\displaystyle{\frac{i}{4\kappa}}
  \left(  {\bar \phi}^b L_b^{\ a}  - L^a_{\ b} \phi^b  \right)     
  + ({\cal D}'_\t)^b \left( L_b^{\ a} - \delta_b^{\ a} \right) 
  - \left( L^a_{\ b} - \delta^a_{\ b} \right) ({{\overline {\cal D}}'_\t})^\b 
   \nn     
& {} & \qquad = 
\displaystyle{\frac{1}{4 \kappa g}} \Biggl( 
L^a_{\ b} { \{  \nabla^\alpha, W_\alpha \}'}^b  
+  { \{  {\overline \nabla}_{\dot \alpha}, {\overline W}^{\dot \alpha} \}'}^b L_b^{\ a}   \Biggr) 
+ {\tilde \eta}^b L_b^{\ a}      \ , 
\ena
as shown in (\ref{eq:Langevin-A-eq4}). Although we have worked in the stochastic action principle for the one-loop renormalization procedure, 
this form of the superfield Langevin equation in the stochastic BFM  may be useful to evaluate anomalies. The analysis will be reported elsewhere.

%%%%%%%%%%%%%%%%%%%%%%%%%%%%%%%%%%%%%%%%%%%%%%%%%%%%%%%%%%%%%%%%%%%%%%%%%%%%%%
%\noindent
%{\bf Acknowledgements}
%%%%%%%%%%%%%%%%%%%%%%%%%%%%%%%%%%%%%%%%%%%%%%%%%%%%%%%%%%%%%%%%%%%%%%%%%%%%%%

\bigskip

\noindent
{\bf Acknowledgements}

The author would like to thank M. B. Halpern and H. Suzuki for valuable comments.

%\newpage
%%%%%%%%%%%%%%%%%%%%%%%%%%%%%%%%%%%%%%%%%%%%%%%%%%%%%%%%%%%%%%%%%%%%%%%%%%%%%%
\appendix

%%%%%%%%%%%%%%%%%%%%%%%%%%%%%%%%%%%%%%%%%%%%%%%%%%%%%%%%%%%%%%%%%%%%%%%%%%%%%%
%\section{Conventions for the BFM in SQM}\label{sec:A}
%%%%%%%%%%%%%%%%%%%%%%%%%%%%%%%%%%%%%%%%%%%%%%%%%%%%%%%%%%%%%%%%%%%%%%%%%%%%%%

\section{Conventions in the stochastic BFM}\label{sec:A}

A backgound superfield $\Omega$, $\Omega^\dagger$ and a quantum superfield $V$ are introduced by the following definition.\cite{GSR}
\bea
\label{eq:background-quantum-splitt-A-eq1}
e^{2g{\hat V}} 
\equiv 
e^{g\Omega}e^{2gV} e^{g\Omega^\dagger} \ .
\ena
Here we have denoted the original vector superfield as ${\hat V}$. In this note except \S 2 and \S 3, \lq\lq ${\hat {\cal O}}$ \rq\rq indicates that the quantity ${\cal O}$ such as a covariant derivative is evaluated with respect to the original vector superfield ${\hat V}$ or the quantity is defined in relation to the original vector superfield. We note that the quantum fluctuation $V$ is a vector superfield, while the background fields, $\Omega$ and $\Omega^\dagger$, are not. 
There are two types of local gauge transformations. One is the background type 
\bea
\label{eq:background-local-gauge-transf-A-eq1}
e^{2g{\hat V}} 
& \rightarrow & 
(e^{-ig{\hat \Sigma}^\dagger}e^{g\Omega}e^{ig K})
e^{-ig K}e^{2gV} e^{ig K}
(e^{-ig K}e^{g\Omega^\dagger}e^{ig {\hat \Sigma}})                    \ ,   \nn 
e^{ig {\hat \Lambda}} 
& \rightarrow & 
e^{ig {\hat \Lambda}} e^{ig {\hat \Sigma}}    \ ,
\ena
where $K$ is a vector superfield, $K^\dagger = K$. 
The superfields ${\hat \Sigma}$ and ${\hat \Sigma}^\dagger$ are chiral and anti-chiral, ${\overline D}_{\dot \a}{\hat \Sigma} =  D_\a {\hat \Sigma}^\dagger = 0$, respectively. The background vector superfield is defined by $e^{2g{\bf V}} \equiv e^{g\Omega}e^{g\Omega^\dagger}$ which is transformed as 
\bea
\label{eq:background-local-gauge-transf-A-eq2}
e^{2g{\bf V}} 
\rightarrow 
e^{-ig{\hat \Sigma}^\dagger}e^{2g{\bf V}}e^{ig {\hat \Sigma}}         \  ,    
\ena
%. 
under the background local gauge transformation. By using the gauge degrees of freedom of $K$, we can choose the gauge $\Omega = \Omega^\dagger = {\bf V}$.\cite{GGRS} In this gauge, the residual background local gauge transformation is (\ref{eq:background-local-gauge-transf-A-eq2}). 

The other is the quantum type local gauge transformation, 
\bea
\label{eq:quantum-local-gauge-transf-A-eq1}
e^{2g{\hat V}} 
& \rightarrow &
e^{g\Omega}(e^{- g\Omega}e^{-ig{\hat \Sigma}^\dagger}e^{g\Omega})
e^{2gV} (e^{g\Omega^\dagger}e^{ig{\hat \Sigma}}e^{-g\Omega^\dagger})e^{g\Omega^\dagger}  \ ,   \nn 
e^{ig {\hat \Lambda}} 
& \rightarrow & 
e^{ig {\hat \Lambda}} e^{ig {\hat \Sigma}}    \ .
\ena
The quantum type is generated by the superfield 
$\Sigma \equiv e^{g\Omega^\dagger} {\hat \Sigma} e^{-g\Omega^\dagger}$ and its hermitian conjugate $\Sigma^\dagger$. They are background covariantly chiral and background covariantly anti-chiral, satisfying 
\bea
\label{eq:background-chiral-cond-A-eq1}
{\overline {\cal D}}_{\dot \alpha}  \Sigma
\equiv 
[ {\overline {\cal D}}_{\dot \alpha},  \Sigma ] 
= 0 \ ,  \quad
{\cal D}_\alpha  \Sigma^\dagger 
\equiv 
[ {\cal D}_\alpha, \Sigma^\dagger ]
= 0    \   .
\ena
Here the background covariant spinor derivatives are defined by 
\bea
\label{eq:background-covariant-derivative-A-eq1}
{\cal D}_\alpha 
 \equiv 
e^{- g\Omega} D_\alpha e^{g\Omega}  
= 
e^{- gL_\Omega} D_\alpha  , \quad 
{\overline {\cal D}}_{\dot \alpha} 
 \equiv 
e^{g\Omega^\dagger} {\overline D}_{\dot \alpha} e^{-g\Omega^\dagger} = 
e^{gL_\Omega^\dagger} {\overline D}_{\dot \alpha}  , \
\ena

The background covariant spinor derivatives are transformed as 
\bea
\label{eq:background-covariant-derivative-A-eq2}
 {\cal D}_\alpha  
\rightarrow 
e^{-igK}
 {\cal D}_\alpha e^{igK} , \quad 
{\overline {\cal D}}_{\dot \alpha} 
\rightarrow 
e^{-igK}
 {\overline {\cal D}}_{\dot \alpha} e^{igK}  
 \ ,
\ena
under the background local gauge transformation. 
We also introduce a background covariant space-time derivative ${\cal D}_m$ 
\bea
\label{eq:background-covariant-derivative-A-eq2'}
\Big\{ {\cal D}_\alpha,\ {\overline {\cal D}}_{\dot \alpha} \Big\} 
= -2i\sigma^m_{\alpha{\dot \alpha}} {\cal D}_m   \ . 
\ena

The superfield strength, $W_\a$, is expanded with respect to the quantum field $V$ in the following way. 
The original covariant derivative, 
${\hat {\cal D}}_\alpha \equiv e^{- 2g{\hat V}} D_\alpha e^{2g{\hat V}} = e^{- 2gL_{\hat V}} D_\alpha$, satisfies a covariant commutation relation with ${\overline D}_{\dot \alpha}$
\bea
\label{eq:background-covariant-derivative-A-eq3}
\Big\{ {\hat {\cal D}}_\alpha,\ {\overline D}_{\dot \alpha} \Big\} 
= -2i\sigma^m_{\alpha{\dot \alpha}} {\hat {\cal D}}_m   \ ,
\ena
where 
$
\Big\{ e^{- 2g{\hat V}} ( D_\alpha e^{2g{\hat V}} ),\  {\overline D}_{\dot \alpha}
 \Big\} 
 \equiv 2\sigma^m_{\alpha{\dot \alpha}} {\hat \Gamma}_m 
$ and 
$ 
{\hat {\cal D}}_m \equiv \pa_m +i {\hat \Gamma}_m 
$. 
It is more convenient to introduce a chiral representation by a similarity transformation, 
\bea
\label{eq:background-covariant-derivative-A-eq4}
( \nabla_\alpha,\  {\overline {\cal D}}_{\dot \alpha},\  \nabla_m   ) 
\equiv 
( e^{g L_\Omega^\dagger} {\hat {\cal D}}_\alpha,\ e^{g L_\Omega^\dagger} {\overline D}_{\dot \alpha},\ e^{g L_\Omega^\dagger} {\hat {\cal D}}_m 
) 
. \ 
\ena
\lq\lq $\nabla$ \rq\rq denotes that the covariant derivative includes the quantum vector superfield $V$. For the other chiral representation of covariant derivatives 
${\hat {\overline {\cal D}}}_{\dot \alpha} \equiv e^{ 2g{\hat V}} {\overline D}_{\dot \alpha} e^{-2g{\hat V}} = e^{ 2gL_{\hat V}} {\overline D}_{\dot \alpha}$ and $D_\alpha$, which satisfy 
\bea
\label{eq:background-covariant-derivative-A-eq5}
\Big\{ {\hat {\overline {\cal D}^{\dot \alpha}}} ,\ D^\alpha \Big\}  
= -2i{\overline \sigma}_m^{{\dot \alpha}\alpha} {\hat {\overline {\cal D}}^m}   \ ,
\ena
we introduce, 
\bea
\label{eq:background-covariant-derivative-A-eq6}
( {\cal D}^\alpha,\  {\overline \nabla}^{\dot \alpha},\  {\overline \nabla}_m   ) 
\equiv 
( e^{-g L_\Omega} D^\alpha,\ e^{-g L_\Omega} {\hat {\overline {\cal D}^{\dot \alpha}}},\ e^{-g L_\Omega} {\hat {\overline {\cal D}}_m} 
)         . \ 
\ena
Superfield strengthes and their covariant derivatives are also expressed with these covariant derivatives by the similarity transformation,
\bea
\label{eq:background-field-strength-A-eq1}
( 
 W_\alpha,\quad 
\{ \nabla^\beta,\ W_\beta \} 
)        
& \equiv & 
( 
e^{g L_\Omega^\dagger} {\hat W}_\alpha,\quad  
e^{g L_\Omega^\dagger} \{ {\hat {\cal D}}^\beta,\ {\hat W}_\beta \} 
)       \ , \nn 
(
{\overline W}_{\dot \alpha},\quad  
\{ {\overline \nabla}_{\dot \beta},\ {\overline W}^{\dot \beta} \} 
)                  
& \equiv &
( 
e^{-g L_\Omega} {\hat {\overline W}}_{\dot \alpha},\quad  
e^{-g L_\Omega} \{ {\hat {\overline {\cal D}}}_{\dot \beta},\  {\hat {\overline W}^{\dot \beta}} \} 
)   \ , 
\ena
where 
\bea
\label{eq:background-field-strength-A-eq2}
W_\alpha 
& = & 
-\displaystyle{\frac{1}{8}} 
[ {\overline {\cal D}}_{\dot \beta},\ 
\{ {\overline {\cal D}}^{\dot \beta},\ \nabla_\alpha \} ] \ ,\nn
{\overline W}_{\dot \alpha}
& = &
\displaystyle{\frac{1}{8}} 
[ {\cal D}^\beta,\ \{ {\cal D}_\beta,\ {\overline \nabla}_{\dot \alpha} \} ] 
     \ . \
\ena
 We also note the reality condition. In its original form, it is given by 
$e^{2g L_{\hat V}} \{ {\hat {\cal D}}^\a, {\hat W}_\a \} = \{ {\hat {\overline {\cal D}}}_{\dot \a},\ {\hat {\overline W}}^{\dot \a} \}$.  This reads
\bea
\label{eq:realitiy-cond-A-eq1}
e^{2g L_V} \{ \nabla^\a, W_\a \} = \{ {\overline \nabla}_{\dot \a},\ {\overline W}^{\dot \a} \}         \    , 
\ena
that holds in each order of $V$. 

These formulaue are expressed by the background vector superfield ${\bf V}$ defined by $e^{2g{\bf V}} \equiv e^{g\Omega}e^{g\Omega^\dagger}$ and the background covariant derivatives
\bea
\label{eq:background-covariant-derivative-A-eq7}
{\bf D}_\a  
& \equiv & e^{-2g {\bf V}} D_\a e^{2g {\bf V}}  \ , \nn
{\overline {\bf D}}_{\dot \a}  
& \equiv & e^{2g {\bf V}} {\overline D_{\dot \a}} e^{-2g {\bf V}}  \ .
\ena
The corresponding superfield strengthes, ${\bf W}_\a$ and ${\overline {\bf W}}_{\dot \a}$, are also introduced by replacing $D_\a$ and ${\overline D}_{\dot \a}$ to ${\bf D}$ and ${\overline {\bf D}}_{\dot \a}$, respectively, in their original definition (\ref{eq:Gluino-field}). In this note, we use the relations in the gauge 
$\Omega = \Omega^\dagger = {\bf V}$ 
\bea
\label{eq:background-vector-superfield-eq1}
{\bf W}_\a 
= e^{-g L_{\Omega^\dagger}} W^{(0)}_\a \ , \quad
{\overline {\bf W}}_{\dot \a} 
= e^{g L_{\Omega}} {\overline W}^{(0)}_{\dot \a} \ , {\rm e}.{\rm t}.{\rm c}. 
\ena
Here 
$W^{(0)}_\a \equiv  W_\a \Big|_{V=0}$ 
and 
${\overline W}^{(0)}_{\dot \a} \equiv  {\overline W}_{\dot \a} \Big|_{V=0}$.

After the background local gauge invariant gauge fixing procedure, the stochastic BFM respects the invariance under the stochastic time $independent$ backgound local gauge transformation for which the transformation parameters are stochastic time $independent$, $\displaystyle{\frac{d}{dt}}{\hat \Sigma} = \displaystyle{\frac{d}{dt}} {\hat \Sigma}^\dagger = 0$ and $\displaystyle{\frac{d}{dt}}K = 0$. This is because the stochastic gauge fixing procedure $breaks$ the invariance under the stochastic time $dependent$ local gauge transformation. For the stochastic time $independent$ background local gauge transformation, we introduce the background covariant derivative with respect to the stochastic time 
\bea
\label{eq:background-covariant-time-derivarive-A-eq1}
{\cal D}_\t 
& = & 
 e^{-g \Omega} \displaystyle{\frac{d}{d\t}} e^{g \Omega}    \ , \nn
{\overline {\cal D}}_\t
& = & 
 e^{g\Omega^\dagger} \displaystyle{\frac{d}{d\t}} e^{-g\Omega^\dagger}  \ , 
\ena
which are transformed as 
$
 {\cal D}_\t  
\rightarrow 
e^{-igK}
 {\cal D}_\t e^{igK} , \quad 
{\overline {\cal D}}_\t 
\rightarrow 
e^{-igK}
 {\overline {\cal D}}_\t e^{igK}  
 \ .
$
Corresponding to this definition, the stochastic time independence is expressed as the background $covariant$ stochastic time independence, which means that the transformation parameters satisfy
\bea
\label{eq:background-covariant-time-derivarive-A-eq2}
{\overline {\cal D}}_\t \Sigma = {\cal D}_\t \Sigma^\dagger = 0  \ . 
\ena

The covariant derivatives (\ref{eq:background-covariant-time-derivarive-A-eq1}) are also expressed with the background vector superfield as 
$ 
{\overline {\bf D}}_\t \equiv e^{2g {\bf V}}\displaystyle{\frac{d}{d\t}}e^{-2g {\bf V}}\ . 
$
This reads
\bea
\label{eq:background-covariant-time-derivarive-A-eq3}
{\overline {\bf D}}_\t 
= e^{gL_\Omega} {\overline {\cal D}}_\t 
= \displaystyle{\frac{d}{d\t}} +  e^{gL_\Omega} ( {\overline {\cal D}}'_\t 
- {\cal D}'_\t )            \   .
\ena
Here ${\overline {\cal D}}'_\t 
\equiv  {\overline {\cal D}}_\t - \displaystyle{\frac{d}{d\t}}  $ 
and 
$ {\cal D}'_\t 
\equiv {\cal D}_\t  - \displaystyle{\frac{d}{d\t}} $. 
We note that this simply indicates that the derivative with respect to the stochastic time must appear in the effective stochastic action in the combination 
$
{\overline {\cal D}}'_\t - {\cal D}'_\t 
$.

%\newpage
%%%%%%%%%%%%%%%%%%%%%%%%%%%%%%%%%%%%%%%%%%%%%%%%%%%%%%%%%%%%%%%%%%%%%%%%%%%%%%%
%\section{Derivation of the Langevin equation in the background field 
%method}\label{sec:B}
%%%%%%%%%%%%%%%%%%%%%%%%%%%%%%%%%%%%%%%%%%%%%%%%%%%%%%%%%%%%%%%%%%%%%%%%%%%%%%%

\section{Derivation of the Langevin equation in the stochastic BFM}\label{sec:B}

In this note, we work in the stochastic action principle for the perturbative analysis and do not use the Langevin equation, explicitly. In the stochastic BFM, the Langevin equation is useful for the evaluation of anomalies. For further application of the stochastic BFM, 
we derive the Langevin equation in this appendix. 
we consider the continuum limit of the Langevin equation (\ref{eq:Langevin-eq2}) by taking the limit $\Delta \t \rightarrow 0$. The noise superfield $\Delta w$ is replaced to $\eta$. The correlation of $\eta$ is defined in (\ref{eq:continuum-noise-correlation-eq1}). 
We assume that the background fields depend on the stochastic time. 
%%%%%%%%%%%%%%%%%%%%%%%%%%%%%%%%%%%%%%%%%%%%%%%%%%%%%%%%%%%%%%%%%%%%%%%%%%%%%%%
%Under the definition, after taking the continuum limit of the Langevin equation% (\ref{eq:Langevin-eq1}), the l.h.s. of it reads, 
%
%\bea
%\label{eq:Langevin-A-eq1}
%& {} & 
%%\displaystyle{\frac{1}{2g}} 
%( \displaystyle{\frac{d}{d\t}} e^{2{\hat V}} )e^{-2{\hat V}}     \nn
%& {} & =
%\displaystyle{\frac{1}{2g}} 
% e^{L_\Omega} \Big( 
% \displaystyle{\frac{e^{2L_V} -1 }{L_V}} \displaystyle{\frac{d}{d\t}}V  
% + e^{2g L_V} ( ( \displaystyle{\frac{d}{d\t}} e^{g\Omega^\dagger} ) e^{-g\Omeg%a^\dagger} ) 
% - ( \displaystyle{\frac{d}{d \t}} e^{-g \Omega} ) e^{g \Omega}
%& \Big)     \ .
%\ena
%%%%%%%%%%%%%%%%%%%%%%%%%%%%%%%%%%%%%%%%%%%%%%%%%%%%%%%%%%%%%%%%%%%%%%%%%%%%%%

The auxiliary chiral and anti-chiral superfields for the gauge fixing functions , ${\hat \Phi}$ and ${\hat {\bar \Phi}}$ are now the background covariantly chiral and the background covarianly anti-chiral. Namely, we have redefined 
$
%\label{eq:background-invariant-gauge-fixing-A-eq1}
 \phi \equiv e^{g L_{\Omega^\dagger}} {\hat \Phi}  \  
$ 
and 
$  
 {\bar \phi} \equiv e^{- g L_\Omega} {\hat {\bar \Phi}}     \ 
$
in (\ref{eq:background-invariant-gauge-fixing-eq1}). 
${\hat {\bar \Phi}} = {\hat \Phi}^\dagger$ implies ${\bar \phi} = \phi^\dagger$.
These superfields satisfy, 
$
%\label{eq:background-invariant-gauge-fixing-A-eq2}
{\overline {\cal D}}_{\dot \a} \phi = {\cal D}_\a {\bar \phi} = 0    \ .
$
By the definitions of the background covariant derivatives and their chiral representation, we obtain
\bea 
\label{eq:Langevin-A-eq2} 
& {} & 
\displaystyle{\frac{1}{2g}} \Big\{ 
\displaystyle{\frac{e^{2gL_V} -1 }{L_V} } \displaystyle{\frac{d}{d\t}} V 
  + e^{2g L_V} \Big( 
  ( \displaystyle{\frac{d}{d\t}} e^{g\Omega^\dagger} ) e^{-g\Omega^\dagger} 
  \Big) 
 - ( \displaystyle{\frac{d}{d\t}} e^{-g \Omega} ) e^{g \Omega}
 \Big\}    \nn
& {} & =  
+ \displaystyle{\frac{i}{4\kappa}}
  (  {\bar \phi}  - e^{2g L_V} \phi  )          
+ \displaystyle{\frac{1}{4 \kappa g}} \Big(  
e^{2gL_V} \{  \nabla^\alpha, W_\alpha \}  
+  \{  {\overline \nabla}_{\dot \alpha}, {\overline W}^{\dot \alpha} \}  
 \Big) 
+ e^{-g L_\Omega} \eta      \ . 
\ena
Under the stochastic time $independent$ background local gauge transformation, the l.h.s. is transformed as 
$
( \displaystyle{\frac{d}{d\t}} e^{2gV} )e^{-2gV} 
\rightarrow 
e^{-ig K}( \displaystyle{\frac{d}{d\t}} e^{2gV} )e^{-2gV} e^{igK}  \ ,
$
thereby ensuring the covariance of the Langevin equation under the stochastic time $independent$ background local gauge transformation. 

In a standard analysis of SYM$_4$ in SQM, the Zwanziger's gauge fixing functions have been chosen as 
$
{\hat \Phi} 
 =  i {\xi\over 4}{\overline D}^2 D^2 {\hat V}    \ 
$
and 
$
{\hat {\bar \Phi}} 
 = - i {\xi\over 4} D^2 {\overline D}^2 {\hat V}  \  \
$
in (\ref{eq:stochastic-gauge-fixing-eq1}). 
In the stochastic BFM , 
for the covariance of the gauge fixing functions under the background local gauge transformation, they are determined on dimensional ground as 
$
%\label{eq:background-invariant-gauge-fixing-A-eq4}
\phi
= 
i {\xi\over 4}
{\overline {\cal D}}^2 {\cal D}^2 V \ 
$
and 
$
{\bar \phi}
= 
- i {\xi\over 4}
{\cal D}^2 {\overline {\cal D}}^2 V  \
$
in (\ref{eq:background-stochastic-gauge-fixing-eq1}). 
This choice satisfies the background chiral and anti-chiral conditions on the gauge fixing functions and conincides with the original gauge fixing functions in the weak background field limit. 
For the quantum vector superfield $V$, (\ref{eq:Langevin-A-eq3}) reads 
\bea
\label{eq:Langevin-A-eq3} 
& {} & 
 \Big\{ 
 \displaystyle{\frac{d}{d\t}} V 
  + \displaystyle{\frac{1}{2g}} L^a_{\ b} \Big( 
  ( \displaystyle{\frac{d}{d\t}} e^{g\Omega^\dagger} ) e^{-g\Omega^\dagger} 
  \Big)^b 
 - \displaystyle{\frac{1}{2g}} \Big( 
 ( \displaystyle{\frac{d}{d\t}} e^{-g \Omega} ) e^{g \Omega} 
 \Big)^b L_b^{\ a}
 \Big\}     
-   
\displaystyle{\frac{i}{4\kappa}}
  (  {\bar \phi}^b L_b^{\ a}  - L^a_{\ b} \phi^b  )      \nn     
& {} & = 
\displaystyle{\frac{1}{4 \kappa g}} \Big(  
L^a_{\ b} \{  \nabla^\alpha, W_\alpha \}^b  
+  \{  {\overline \nabla}_{\dot \alpha}, {\overline W}^{\dot \alpha} \}^b L_b^{\ a}   \Big) 
+ (e^{-g L_\Omega} \eta )^b L_b^{\ a}      \ . 
\ena
The stochastic action in the stochastic BFM can be directly obtained from the integral representation of this expression. 

We note that one might suspect that the noise superfield ${\tilde \eta} \equiv e^{-g L_\Omega} \eta$ in (\ref{eq:Langevin-A-eq3}) depends on the background field $\Omega$ and $\Omega^\dagger$, however, this is not the case. In fact, by definition, the noise superfield, ${\tilde \eta}$, satisfies the correlation (\ref{eq:continuum-noise-correlation-eq1}). The hermiticity condition on the noise superfield, 
$ 
\eta^\dagger = e^{-2g L_{\hat V}} \eta 
$, is reduced to 
\bea
\label{eq:background-nose-A-eq1}
{\tilde \eta}^\dagger = e^{-2g L_V} {\tilde \eta}   \ , 
\ena
for ${\tilde \eta}$. From (\ref{eq:background-nose-A-eq1}), ${\tilde \eta}^\dagger$ also satisfies the correlation (\ref{eq:continuum-noise-correlation-eq1}). The correlation between ${\tilde \eta}$ and ${\tilde \eta}^\dagger$ is given by 
\bea
\label{eq:background-nose-A-eq2}
\langle {\tilde \eta}_{ij} (\t, z){\tilde \eta}^\dagger (\t', z') \rangle_{{\tilde \eta}_\t} 
 =  
2\beta {\rm k} \Big( 
( e^{2gV} )_{il} ( e^{-2gV} )_{kj} - \displaystyle{\frac{1}{N}} \delta_{ij}\delta_{kl} 
\Big) \delta^8 (z-z') \delta ( \t-\t' )     \  .
\ena
Here, the r.h.s. is $not$ the expectation value in It${\bar {\rm o}}$ calculus.  Therefore, ${\tilde \eta}$ has no dependence on the background fields. The components of ${\tilde \eta}^\dagger$ are determined by (\ref{eq:background-nose-A-eq1}). The condition (\ref{eq:background-nose-A-eq1}) means that the noise superfield ${\tilde \eta}$ (also $\eta$) is not a vector superfield. This is the price we have paid for the covariance of the superfield Langevin equation. We emphasize that, although the noise superfield is not a vector superfield, the time evolution of the vector superfield $V$ preserves its hermiticity $V^\dagger = V$. This can be seen as follows. (\ref{eq:background-nose-A-eq1}) reads 
\bea
\label{eq:background-nose-A-eq3}
{\tilde \eta} = \Big\{ 
1 + {\rm tanh}(gL_V) \Big\} \displaystyle{\frac{1}{2}} ( {\tilde \eta} + {\tilde \eta}^\dagger )           \ .
\ena
The second term represents the imaginary part of the complex noise superfield ${\tilde \eta}$. On the other hand, the time evolution equation of the vector superfield $V$ in (\ref{eq:Langevin-A-eq3}) is described by the collective noise 
${\tilde \Xi}_{\tilde \eta} \equiv {\tilde \eta}^b L_b^{\ a}$. In the matrix representation, as given in (\ref{eq:Langevin-eq1'}) for the time evolution of ${\hat V}$ before taking the continuum limit, this is expressed as 
\bea
\label{eq:background-nose-A-eq4}
{\tilde \Xi}_{\tilde \eta} 
& = &  
\displaystyle{\frac{2gL_V}{e^{2gL_V} - 1}} {\tilde \eta}     \  , \nn
& = & 
\Big\{ 
{\rm coth}(gL_V) - {\rm tanh}(gL_V)
\Big\}gL_V \displaystyle{\frac{1}{2}} ( {\tilde \eta} + {\tilde \eta}^\dagger )     \ . 
\ena
As is clear from the expression, the collective noise is a vector superfield, 
$
{\tilde \Xi}_{\tilde \eta}^\dagger = {\tilde \Xi}_{\tilde \eta} 
$. 
This ensures that the time evolution of the vector superfield $V$ preserves its hermiticity $V^\dagger = V$. The relations we have explained for the noise superfields in the stochastic BFM also hold in the standard SQM approach.\cite{Nakazawa1}\cite{Nakazawa2} 
  
In the Langevin equation (\ref{eq:Langevin-A-eq3}), we discard the \lq\lq classical \rq\rq field equations for the background fields. It is given by 
\bea
\label{eq:classical-equations-motion-A-eq1}
\Big( 
( \displaystyle\frac{d}{d\t}e^{g\Omega^\dagger} )e^{-g\Omega^\dagger} 
+ e^{-g\Omega} ( \displaystyle\frac{d}{d\t}e^{g\Omega} ) 
\Big) 
- \frac{1}{2\kappa} \Big( \{ {\cal D}^\a,\ W_\a^{(0)} \} 
+  \{ {\overline {\cal D}}_{\dot \a},\ {\overline W}^{{\dot \a} (0)} \}  
\Big)  = 0    \ .
\ena
The classical backgound field equation is written by $\Omega$ and $\Omega^\dagger$. By the definitions of the covariant derivatives with respect to ${\bf V}$, it reads 
\bea
\label{eq:background-field-equation-A-eq1}
{\overline {\bf D}}'_\t  
+ \frac{1}{2\kappa} \Big( 
e^{2g {L_{\bf V}}}\{ {\bf D}^\a,\ {\bf W}_\a \} 
+  \{ {\overline {\bf D}}_{\dot \a},\ {\overline {\bf W}}^{\dot \a} 
\} \Big)  = 0    \ .
\ena
Substituting the classical field equation (\ref{eq:classical-equations-motion-A-eq1}) into the Langevin equation (\ref{eq:Langevin-A-eq3}), we obtain 
\bea
\label{eq:Langevin-A-eq4} 
& {} &  
 \displaystyle{\frac{d}{d\t}} V^a     
- 
\displaystyle{\frac{i}{4\kappa}}
  \left(  {\bar \phi}^b L_b^{\ a}  - L^a_{\ b} \phi^b  \right)     
  + ({\cal D}'_\t)^b \left( L_b^{\ a} - \delta_b^{\ a} \right) 
  - \left( L^a_{\ b} - \delta^a_{\ b} \right) ({{\overline {\cal D}}'_\t})^\b 
   \nn     
& {} & \qquad = 
\displaystyle{\frac{1}{4 \kappa g}} \Biggl( 
L^a_{\ b} { \{  \nabla^\alpha, W_\alpha \}'}^b  
+  { \{  {\overline \nabla}_{\dot \alpha}, {\overline W}^{\dot \alpha} \}'}^b L_b^{\ a}   \Biggr) 
+ {\tilde \eta}^b L_b^{\ a}      \ . 
\ena
Here 
$ \{  \nabla^\alpha, W_\alpha \}'
=  \{  \nabla^\alpha, W_\alpha \} - \{  {\cal D}^\alpha, W_\alpha \}  \ ,
$ 
and 
$
\{  {\overline \nabla}_{\dot \alpha}, {\overline W}^{\dot \alpha} \}' 
= \{  {\overline \nabla}_{\dot \alpha}, {\overline W}^{\dot \alpha} \} 
- \{  {\overline D}_{\dot \alpha}, {\overline W}^{\dot \alpha} \}   \ .
$
This is the basic Langevin equation for the stochastic BFM. In the Feynman gauge $\xi=1$, the expression becomes 
\bea
\label{eq:Langevin-A-eq5} 
& {} & 
\displaystyle{\frac{d}{d\t}} V - 
\displaystyle{\frac{1}{2}} \Big[\  V,\ 
 {\cal D}'_\t + {\overline {\cal D}}'_\t\ \Big]     \nn
& {} & \quad 
=  
\displaystyle{\frac{1}{\kappa}} \left(
{\cal D}^m{\cal D}_m - W^{\alpha (0)}{\cal D}_\a 
 + {\overline W}_{\dot \alpha}^{(0)} {\overline {\cal D}}^{\dot \alpha}  
 \right) V 
 + O ( V^n, n \ge 2 )  + \eta           \ , \nn
& {} & 
O( V^n, n \ge 2  )    \nn 
& {} & \quad 
=  
- \displaystyle{\frac{i}{4\kappa}}g 
[\ V,\ \phi^\dagger + \phi\ ] 
+ \displaystyle{\frac{1}{\kappa}} \Big( 
- g \{\ [\ V,\ {\overline {\cal D}}_{\dot \a}V\ ],\ {\overline W}^{{\dot a} (0)}\ \} 
+ \displaystyle{\frac{g}{4}} \{\ {\overline {\cal D}}_{\dot \a}V,\ {\cal D}^2 {\overline {\cal D}}^{\dot a} V\ \}        \nn
& {} & \qquad 
- \displaystyle{\frac{g}{8}}  {\overline {\cal D}}_{\dot \a} {\cal D}^2 [\ V,\ {\overline {\cal D}}^{\dot \a}\ ]   
+ g [\ V,\ \{\ {\overline {\cal D}}_{\dot \a}V,\ {\overline W}^{{\dot a} (0)}\
   \}\ ] 
+ \displaystyle{\frac{g}{8}}  [\ V,\ {\overline {\cal D}}_{\dot \a} {\cal D}^2 {\overline {\cal D}}^{\dot a} V\ ]    
\Big)      \nn
& {} & \qquad 
- \displaystyle{\frac{g}{6}}[\ V,\ [\ V,\ 
{\cal D}'_\t - {\overline {\cal D}}'_\t 
- \displaystyle{\frac{1}{\kappa}} {\overline {\cal D}}_{\dot \a} {\overline W}^{{\dot a} (0)}\ ]\ ]  + O ( V^{n'}, n' \ge 3  ) 
- g [\ V,\ {\tilde \eta}\ ]       \ . 
\ena
Here, for the one-loop analysis, the interaction terms are necessary up to the second order of the quantum fluctuation, $V$. 
In order to derive the expression, we have used the reality condition and the following relations
\bea
\label{eq:background-formulae-A-eq1}
{\cal D}^\a{\overline {\cal D}}^2 {\cal D}_\a - \displaystyle{\frac{1}{2}} ( {\cal D}^2{\overline {\cal D}}^2 + {\overline {\cal D}}^2{\cal D}^2 )   
& = & 
- 8 \Big( 
{\cal D}^m{\cal D}_m + {\overline W}_{\dot \a}^{(0)}{\overline {\cal D}}^{\dot \a} 
+ \displaystyle{\frac{1}{2}} {\cal D}^\a W_\a^{(0)} 
\Big)    \ , \nn
{\overline {\cal D}}_{\dot \a}{\cal D}^2{\overline {\cal D}}^{\dot \a} - \displaystyle{\frac{1}{2}} ( {\cal D}^2{\overline {\cal D}}^2 + {\overline {\cal D}}^2{\cal D}^2 )     
& = & 
-8 \Big( 
{\cal D}^m{\cal D}_m - W^{\a\ (0)} {\cal D}_\a
- \displaystyle{\frac{1}{2}} {\overline {\cal D}}_{\dot \a} {\overline W}^{{\dot \a} (0)}  
\Big)    \ . 
\ena
%

%%%%%%%%%%%%%%%%%%%%%%%%%%%%%%%%%%%%%%%%%%%%%%%%%%%%%%%%%%%%%%%%%%%%%%%%%%%%%%
%\section{References}                           2003.2.10 version 
%%%%%%%%%%%%%%%%%%%%%%%%%%%%%%%%%%%%%%%%%%%%%%%%%%%%%%%%%%%%%%%%%%%%%%%%%%%%%%
%

\end{document}